\documentclass[twocolumn]{aastex631}
\usepackage{newtxtext, newtxmath}
\usepackage{natbib}
\usepackage{graphicx}
\usepackage{amsmath}
\usepackage{amssymb}
\usepackage{tabularx}

\defcitealias{Johnson2021}{J21}

\newcommand{\modelname}[2]{\ensuremath{\eta\text{#1-}y#2}}

\newcommand{\msun}{\ensuremath{\text{M}_\odot}}
\newcommand{\grad}[1]{\ensuremath{\nabla\text{[#1/H]}}}
\newcommand{\ycc}[1]{\ensuremath{y_\text{#1}^\text{CC}}}
\newcommand{\yia}[1]{\ensuremath{y_\text{#1}^\text{Ia}}}
\newcommand{\onh}[1]{\ensuremath{\text{[#1/H]}}}
\newcommand{\vice}{{\sc VICE}}
\newcommand{\astronn}{\textsc{AstroNN}}
\newcommand{\gaia}{\textit{Gaia}}
\newcommand{\ddfrac}[2]{\frac{%
    \displaystyle{#1}
}{%
    \displaystyle{#2}
}}

\newcommand{\citepossessivepar}[1]{\citeauthor{#1}'s \citeyearpar{#1}}

\newcommand{\refp}[1]{(\ref{#1})}

\newcommand{\carnegieaffil}{%
Carnegie Science Observatories, %
813 Santa Barbara St., Pasadena, CA 91101, USA}
\newcommand{\osuaffil}{%
Department of Astronomy, The Ohio State University, %
140 W. 18th Ave., Columbus, OH 43210, USA}
\newcommand{\ccappaffil}{%
Center for Cosmology \& Astroparticle Physics (CCAPP), The Ohio State University, %
191 W. Woodruff Ave., Columbus, OH 43210, USA}
\newcommand{\uchileaffil}{%
Departamento de Astronom\'{i}a, Universidad de Chile, Camino El Observatorio 1515, Santiago, Chile}
\newcommand{\cuaffil}{%
Center for Astrophysics \& Space Astronomy, Department of Astrophysical and Planetary Sciences, University of Colorado, %
389 UCB, Boulder, CO 80309, USA}
\newcommand{\vandyaffil}{%
Department of Physics \& Astronomy, Vanderbilt University, %
2301 Vanderbilt Pl., Nashville, TN 37235, USA}
\newcommand{\durhamceaaffil}{%
Centre for Extragalactic Astronomy, Department of Physics, Durham University, %
South Road, Durham DH1 3LE, UK}
\newcommand{\durhamiccaffil}{%
Institute for Computational Cosmology, Department of Physics, Durham University, %
South Road, Durham DH1 3LE, UK}

\newcommand{\uvicaffil}{%
Department of Physics \& Astronomy, University of Victoria, %
3800 Finnerty Rd., Victoria, BC V8P 5C2, Canada}

\newcommand{\amnhaffil}{%
American Museum of Natural History, %
200 Central Park West, New York, NY 10024, USA}

\begin{document}

\title{The Milky Way Radial Metallicity Gradient as an Equilibrium Phenomenon: Why Old Stars are Metal-Rich}
\shorttitle{Chemical Equilibrium in the Galactic Disk}

\author[0000-0002-6534-8783]{James W. Johnson}
\affiliation{\carnegieaffil}
\affiliation{\osuaffil}
\affiliation{\ccappaffil}

\author[0000-0001-7775-7261]{David H. Weinberg}
\affiliation{\osuaffil}
\affiliation{\ccappaffil}

\author[0000-0003-4218-3944]{Guillermo A. Blanc}
\affiliation{\carnegieaffil}
\affiliation{\uchileaffil}

\author[0000-0002-7846-9787]{Ana Bonaca}
\affiliation{\carnegieaffil}

\author[0000-0002-8459-5413]{Gwen C. Rudie}
\affiliation{\carnegieaffil}

\author[0000-0003-4769-3273]{Yuxi (Lucy) Lu}
%%% \affiliation{\columbiaaffil}
\affiliation{\osuaffil}
\affiliation{\amnhaffil}

\author[0000-0002-7187-8561]{Bronwyn Reichardt Chu}
\affiliation{\durhamceaaffil}
\affiliation{\durhamiccaffil}

\author[0000-0001-9345-9977]{Emily J. Griffith}
\altaffiliation{NSF Astronomy and Astrophysics Postdoctoral Fellow}
\affiliation{\cuaffil}

\author[0000-0001-8208-9755]{Tawny Sit}
\affiliation{\osuaffil}
\affiliation{\ccappaffil}

\author[0000-0001-7258-1834]{Jennifer A. Johnson}
\affiliation{\osuaffil}
\affiliation{\ccappaffil}

\author[0000-0003-3781-0747]{Liam O. Dubay}
\affiliation{\osuaffil}
\affiliation{\ccappaffil}

\author[0000-0003-4912-5157]{Miqaela K. Weller}
\affiliation{\osuaffil}
\affiliation{\ccappaffil}

\author[0009-0008-8903-160X]{Daniel A. Boyea}
\affiliation{\uvicaffil}

\author[0000-0001-5838-5212]{Jonathan C. Bird}
\affiliation{\vandyaffil}

% \author{Collaborators}

%%%%%%

\shortauthors{J.W. Johnson et al.}
\correspondingauthor{James W. Johnson}
\email{jjohnson10@carnegiescience.edu}

\begin{abstract}
Metallicities of both gas and stars decline toward large radii in spiral galaxies, a trend known as the radial metallicity gradient.
We quantify the evolution of the metallicity gradient in the Milky Way as traced by APOGEE red giants with age estimates from machine learning algorithms.
Stars up to ages of $\sim$9 Gyr follow a similar relation between metallicity and Galactocentric radius.
This constancy challenges current models of Galactic chemical evolution, which typically predict lower metallicities for older stellar populations.
Our results favor an {\it equilibrium scenario}, in which the gas-phase gradient reaches a nearly constant normalization early in the disk lifetime.
Using a fiducial choice of parameters, we demonstrate that one possible origin of this behavior is an outflow that more readily ejects gas from the interstellar medium with increasing Galactocentric radius.
A direct effect of the outflow is that baryons do not remain in the interstellar medium for long, which causes the ratio of star formation to accretion, $\dot{\Sigma}_\star / \dot{\Sigma}_\text{in}$, to quickly become constant.
This ratio is closely related to the local equilibrium metallicity, since its numerator and denominator set the rates of metal production by stars and hydrogen gained through accretion, respectively.
Building in a merger event results in a perturbation that evolves back toward the equilibrium state on $\sim$Gyr timescales.
Under the equilibrium scenario, the radial metallicity gradient is not a consequence of the inside-out growth of the disk but instead reflects a trend of declining $\dot{\Sigma}_\star / \dot{\Sigma}_\text{in}$ with increasing Galactocentric radius.
\end{abstract}


\section{Introduction}
\label{sec:intro}

Spiral galaxies are metal-rich at small radii and metal-poor at large radii.
The metal mass fraction $Z$ in both gas \citep[e.g.,][]{Wyse1989, Zaritsky1992} and stars \citep[e.g.,][]{Chen2003, Daflon2004, Cheng2012} declines exponentially with Galactocentric radius (more commonly expressed as a linear relation in $\log Z$; e.g., with [O/H] or [Fe/H]), a correlation known as the radial metallicity gradient.
Thanks to the success of spectroscopic surveys, measurements are now available for thousands of external galaxies out to increasingly high redshift \citep[e.g.,][]{Maiolino2019, Sanchez2020}.
First quantified through nebular emission lines in HII regions \citep{Aller1942, Searle1971, Shaver1983}, the presence and ubiquity of gradients are often interpreted as evidence of ``inside-out'' disk growth, in which the inner disk assembles first and the outskirts follow suit on longer timescales \citep[e.g.,][]{Matteucci1989, White1991, Kauffmann1996, Bird2013}.
In this view, the outer regions of spirals have not reached high metallicity because they have not had enough time to do so.
In this paper, we propose an alternative view, namely that metallicity evolves toward an equilibrium state \citep{Finlator2008, Andrews2017, Weinberg2017b}, and this equilibrium abundance declines with Galactocentric radius (\citealt{Johnson2021}; hereafter \citetalias{Johnson2021}).
\par
A common approach to constrain the evolution of the metallicity gradient in the Milky Way (MW) is to analyze mono-age stellar populations.
This methodology is based on the expectation that stars inherit the chemical composition of the local interstellar medium (ISM) when they form (to within $\lesssim$$0.02 - 0.03$ dex; \citealp{DeSilva2006, Bovy2016a, Liu2016b, Casamiquela2020}).
Application of this procedure is limited to the MW where large samples of resolved stars are feasible.
Despite the wealth of observationally accessible targets, the evolution in the ISM metallicity gradient is only weakly constrained.
One source of uncertainty is radial migration \citep[e.g.,][]{Sellwood2002}, which can carry stars several kpc from their birth radius where their abundances no longer reflect the Galactic region in which they formed.
Another source is the difficulty associated with precision age measurements for stars (see, e.g., the reviews by \citealt{Soderblom2010} and \citealt{Chaplin2013}).
Many investigations thus far have used spectral types that coarsely trace underlying populations at young, intermediate, and old ages, such as OB stars \citep[e.g.,][]{Daflon2004}, Cepheid Variables \citep[e.g.,][]{Andrievsky2004, Luck2006, Luck2011a, Luck2011b, Yong2006}, open clusters \citep[e.g.,][]{Friel1995, Chen2003, Magrini2009}, and planetary nebulae \citep[e.g.,][]{Maciel2003, Henry2010, Stanghellini2010, Magrini2016}.
\par
Precise age measurements for red giants are particularly valuable for constraining the enrichment history of the MW disk, since they are accessible at large distances due to their high luminosities.
Of the current methods for measuring stellar ages, asteroseismology is the most reliable for red giants.
Modeling their photometric variability constrains their masses and evolutionary states \citep[e.g.,][]{DeRidder2009, Bedding2010, Bedding2011, Hekker2017}, which enables an age inference based on the mass-lifetime relation \citep[e.g.,][]{Larson1974, Maeder1989, Padovani1993, Kodama1997, Hurley2000}.
Joint catalogs of seismic ages and metal abundances from spectroscopic surveys \citep[e.g. APOKASC;][]{Pinsonneault2014, Pinsonneault2018, Pinsonneault2024} are therefore highly valuable.
Some authors have trained neural networks on these catalogs in order to predict the asteroseismic data from the spectrum and construct large samples of ages \citep[e.g.,][]{Ness2016, Mackereth2019b, Leung2023, Stone-Martinez2024}.
\par
The combination of asteroseismology and spectroscopy has been a popular choice in recent investigations of the age-abundance structure of the Galactic disk \citep[e.g.,][]{Anders2017, SilvaAguirre2018, Spitoni2019, Lian2020b, Lian2020a, Matsuno2021}.
A result of particular interest to this paper is that old and intermediate-aged stars at fixed Galactocentric radius are not significantly more metal poor than young stars.
\citet{Willett2023} demonstrate that conventional models of Galactic chemical evolution (GCE) underpredict the metallicities of old populations.
\citet{Gallart2024} showed that stellar metallicity does not correlate significantly with age up to $\sim$$10$ Gyr in six different age catalogs \citep[see their Figure 13;][]{Xiang2022, Queiroz2023, Kordopatis2023}.
The relationship between metallicity and Galactic radius also does not correlate significantly with age in the latest catalogs of open clusters and Cepheid variables \citep{Spina2022, daSilva2023}.
Some investigations have found variations in gradient slopes with stellar age \citep[e.g.,][]{Carbajo-Hijarrubia2024}, but the result we highlight here is that these variations occur at roughly constant normalization in the metallicity-radius plane.
% These results are likely related to the surprisingly high metallicities observed in high redshift galaxies with the James Webb Space Telescope (JWST; see discussion in Section \ref{sec:discussion:cosmo-context} below).

\begin{figure*}
\centering
\includegraphics[scale = 0.45]{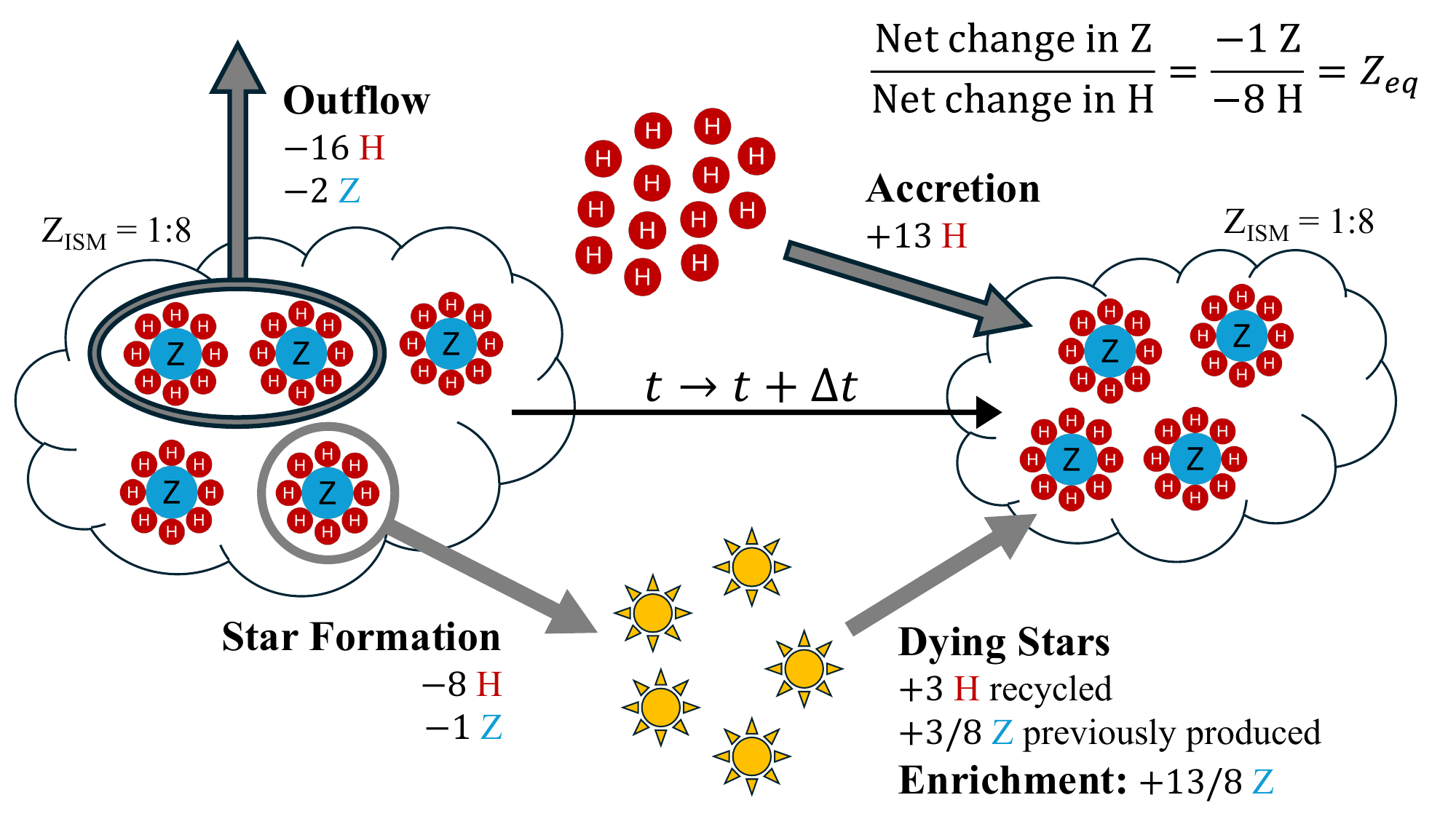}
\caption{
A cartoon of the equilibrium scenario in our models.
We depict the ISM fluid as discrete units, each one unit of metals (Z, blue) and eight units of H (red), so that the composition is countable by eye.
Over the course of one timestep (left to right), a region of the ISM loses one unit to star formation and two to outflows.
Simultaneously, dying stars return three H's and two Z's, while 13 H's are gained through accretion.
The result is a net loss of one fluid element.
The defining feature of equilibrium chemical evolution is that the mixture is unaffected.
}
\label{fig:cartoon}
\end{figure*}

The apparent lack of decline in stellar metallicities across such a broad range of age demands a theoretical explanation.
Such a result is not naturally predicted by ``classical'' GCE models, which imply that metallicities should decline toward old ages because ongoing star formation leads to ongoing metal production (see, e.g., the reviews by \citealt{Tinsley1980} and \citealt{Matteucci2021}).
To provide this explanation, this paper advocates for an {\it equilibrium scenario} as the origin of radial metallicity gradients.
The defining feature of the equilibrium scenario is that the ISM metallicity $Z$ at fixed radius does not evolve significantly with time after the first $\sim$few Gyr of disk evolution.
Instead, $Z$ rapidly evolves toward a local equilibrium abundance $Z_\text{eq}$, which declines with radius in a manner that tracks the observed gradient.
The roots of equilibrium chemical evolution can be traced back to \citet{Larson1974}, who showed that in the presence of ongoing accretion, the ISM metallicity evolves toward an equilibrium at which metal production by stars is balanced by losses to star formation and dilution by metal-poor accreted gas.
\par
Variations in the equilibrium abundance have played a central role in models of the mass-metallicity relation (MZR) for galaxies over the past $\sim$15 years.
High mass galaxies tend to be more metal-rich than their low mass counterparts in terms of both gas \citep[e.g.,][]{Tremonti2004, Andrews2013, Zahid2012, Blanc2019} and stellar populations \citep[e.g.,][]{Gallazzi2005, Kirby2013, Simon2019}.
This trend is often attributed to low mass galaxies more readily ejecting gas in a wind due to their weak gravitational fields \citep[e.g.,][]{Finlator2008, Peeples2011, Zahid2012, Lilly2013, Sanders2021, Chartab2023}.
\citet{Sanchez-Menguiano2024b, Sanchez-Menguiano2024a} recently showed that the scatter in the MZR decreases when quantified in terms of gravitational potential instead of stellar mass, suggesting that metallicity is more sensitive to the local gravity field than just the amount of mass present.
One might therefore expect similar effects if outflows become more efficient as the disk surface density drops with increasing Galactocentric radius.
While the MZR literature invokes variations in chemical equilibrium {\it between} galaxies, our proposed equilibrium scenario invokes these variations {\it within} individual spirals.
\par
Figure \ref{fig:cartoon} shows a cartoon of the equilibrium state that arises in the models in this paper.
Over a small time interval, the ISM undergoes an episode of star formation.
The resultant feedback from massive stars and SNe launches an outflow.
At the same time, the ejected envelopes of dying stars are incorporated into the star forming ISM, and some amount of metal-poor gas is accreted.
In the equilibrium state, these processes lead to net changes in the surface densities of metals and hydrogen that are in the same proportion as the metallicity itself.
The composition of this region of the ISM is unaffected, despite a change in mass.
The equilibrium state is self-stabilizing, since the H gained through accretion fuels star formation, which then produces metals to compensate for the effect of dilution.
The stable equilibrium is therefore noticeable on $\gtrsim$1 Gyr timscales, since stars are expected to re-enrich the ISM quickly following an accretion event \citep[e.g.,][]{Dalcanton2007, Johnson2020}.
This paper is centered around these central themes, as well as how the relative rates of accretion, star formation, and ejection induce variations in the equilibrium metallicity with Galactocentric radius.
\par
% The local equilibrium metallicity is sensitive to the rates of accretion, star formation, and ejection relative to one another during the epoch of thin disk formation.
% The equilibrium state is self-stabilizing, because these processes are closely related.
% For example, if a large amount of H is added to the ISM through accretion, the additional mass will fuel an episode of star formation, which will then supply metals and bring the mixture back toward its original state.
% This process happens quickly following an episode of star formation \citep[e.g.,][]{Dalcanton2007, Johnson2020}.
% As a consequence, the stability is noticeable on $\gtrsim$1 Gyr timescales.
% The relation between the local accretion and ejection rates also plays a key role in setting the local equilibrium metallicity.
% For any choice of GCE parameters, the normalization of the rates of accretion, star formation, and ejection is well constrained by the observed stellar mass of the MW.
% As a consequence, models with strong outflows have high accretion rates in order to make up for the lost mass, which acts to replace metals in the ISM with H.
% These predictions play a key role in holding the ISM metallicity constant in time across the Galactic disk, as indicated by the observations.
% \par
Our GCE models build on the approach of \citetalias{Johnson2021}.
For any study of stellar metallicity gradients, the radial migration of stars is a potentially important process \citep[e.g.,][]{Sellwood2002, Schoenrich2009a, Minchev2013, Minchev2014}.
Following \citetalias{Johnson2021}, we incorporate stellar migration using a recipe calibrated to the predictions of a hydrodynamic simulation of a MW-like galaxy.
Radial gas flows may also influence metal abundances across the Galactic disk, but their treatment is uncertain \citep[e.g.,][]{Lacey1985, Spitoni2011, Bilitewski2012}.
For simplicity, we ignore radial gas flows in this paper; the primary factor that governs metallicity gradients, as in \citetalias{Johnson2021}, is the gas outflow efficiency.
We will examine models incorporating radial gas flows in future work.
\par
In Section \ref{sec:empirical} below, we present evidence in support of the argument that stellar metallicities are age-independent up to at least $\sim$$9 - 10$ Gyr.
In Section \ref{sec:gce}, we construct a set of GCE models that illustrate the equilibrium scenario and contrast them with models in which stellar abundances do not reflect an equilibrium state.
In Section \ref{sec:results}, we compare the predictions of these models with our sample of MW stars and highlight the key differences between parameter choices.
We discuss our models in the context of previous GCE models in the literature as well as the broader scope of galactic astrophysics in Section \ref{sec:discussion}.
We summarize our findings in Section \ref{sec:conclusions}.
In Appendix \ref{sec:leung2023}, we present additional measurements of the Galactic abundance gradient in mono-age populations using a second age catalog as an additional test.
In Appendix \ref{sec:calibration}, we describe how we determine input parameters for our GCE models in detail.

\section{Empirical Age and Metallicity Gradients}
\label{sec:empirical}

\subsection{The Sample}
\label{sec:empirical:apogee}

%fig 1
\begin{figure*}
\centering
\includegraphics[scale = 0.95, trim = 0 -0.15in 0 0]{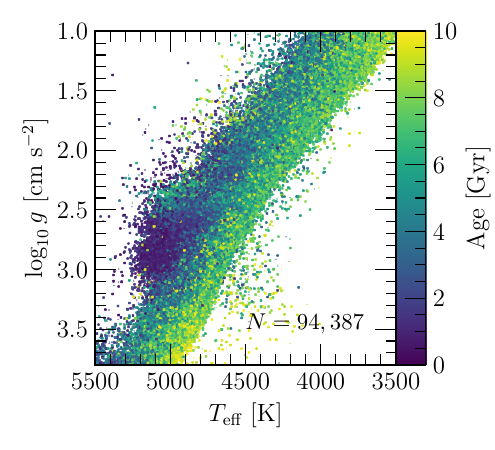}
\includegraphics[scale = 1]{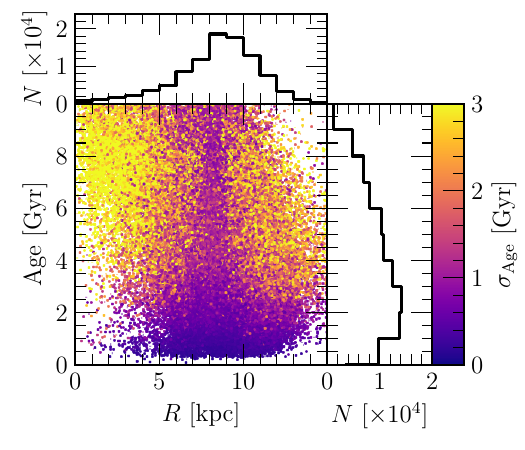}
\caption{
Our sample: the \astronn\ value added catalog \citep{Mackereth2019b} restricted to stars within $\left|z\right| \leq 0.5$ kpc of the disk midplane and the ranges of stellar parameters and Galactocentric radius visualized in these panels (see discussion in Section \ref{sec:empirical:apogee}).
{\bf Left}: The Kiel diagram, color coded by stellar age according to the colorbar.
{\bf Right}: The Galactocentric radii and ages of each star, color coded by the reported age uncertainties according to the colorbar.
Top and right panels show distributions in radius and age, respectively.
{\bf Summary}: By drawing stars from across the red giant branch, we construct a large sample with excellent coverage of the Galactic disk (see discussion in Section \ref{sec:empirical:apogee}).
}
\label{fig:sample}
\end{figure*}

% fig 2
\begin{figure*}
\centering
\includegraphics[scale = 0.95]{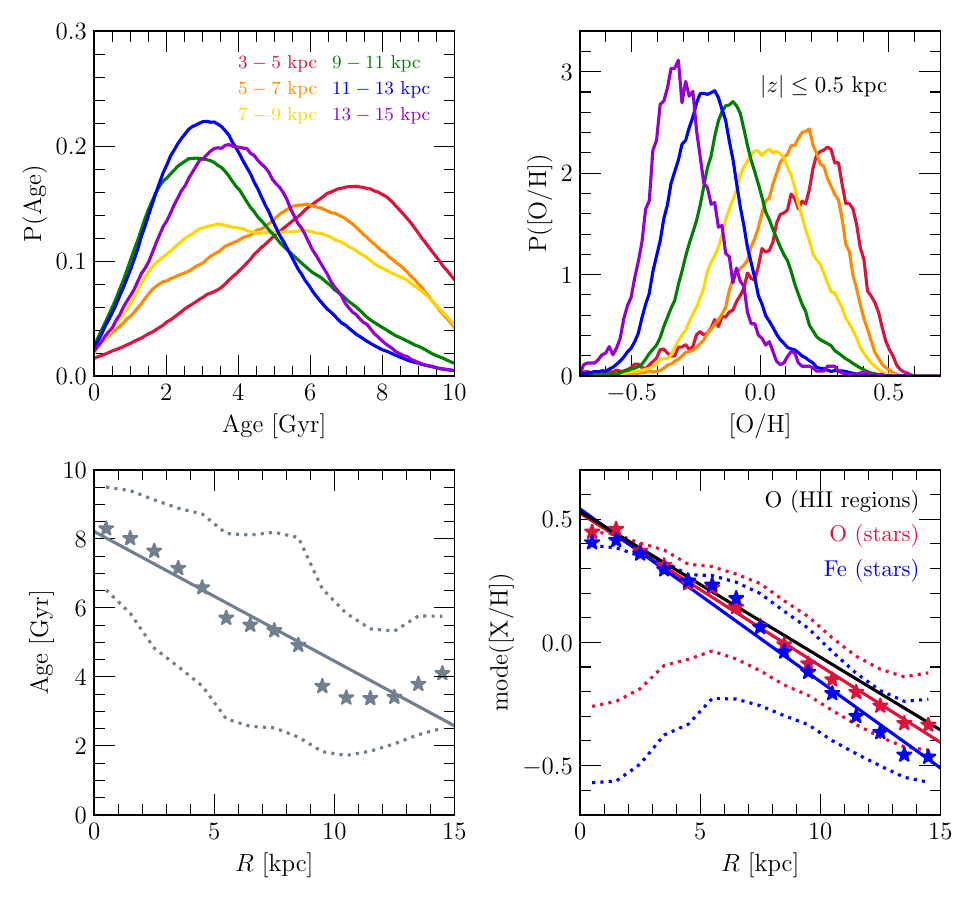}
\caption{
The age-abundance structure of the thin disk as traced by our sample (see Figure \ref{fig:sample} and discussion in Section \ref{sec:empirical:apogee}).
{\bf Top}: Distributions in stellar age (left) and [O/H] (right) in 2-kpc wide bins of Galactocentric radius, color coded according to the legend in the top-left panel.
Each distribution is box-car smoothed with a window width of the median measurement uncertainty in each radial range.
{\bf Bottom}: Median age (left) and the peak of the metallicity distribution (right) in 1-kpc bins of radius (star symbols).
Dotted lines denote the 16th and 84th percentiles of the distribution in each radial range, while solid lines mark the line of best fit to the corresponding summary statistics (parameters of which are given in Table \ref{tab:regressions}).
We additionally plot the gradient in [O/H] measured from Galactic HII regions by \citet[][black]{MendezDelgado2022}.
{\bf Summary}: In agreement with previous work, stellar populations tend to be younger and more metal-poor at large radii.
Distributions in both age and metallicity reverse skewness between the inner and outer disk.
The stellar and ISM metallicity gradients in the Galactic disk are consistent to within 1$\sigma$.
}
\label{fig:gradients}
\end{figure*}

There are many spectroscopic surveys to choose from to characterize the age-abundance structure of the Galactic disk, such as LAMOST \citep{Luo2015}, GALAH~\citep{DeSilva2015, Martell2017}, {\it Gaia}-ESO \citep{Gilmore2012}, and APOGEE\footnote{
    LAMOST: Large sky Area Multi-Object fibre Spectroscopic Telescope
    \\
    GALAH: GALactic Archaeology with Hermes
    \\
    ESO: European Southern Observatory
    \\
    APOGEE: Apache Point Observatory Galaxy Evolution Experiment
} \citep{Majewski2017}.
APOGEE is particularly conducive to this task, since it targets luminous evolved stars accessible at large distances and is less susceptible to dust obscuration with spectra taken at near-IR wavelengths \citep[$\lambda = 1.51 - 1.70~\mu$m;][]{Wilson2019}.
In this paper, we focus on the\ \astronn\ value added catalog\footnote{
    \url{https:/www.sdss.org/dr18/data_access/value-added-catalogs/?vac_id=85}
}
for APOGEE's seventeenth data release \citep[DR17;][]{Abdurrouf2022}.
In making the age measurements for the original value added catalog, \citet{Mackereth2019b} used \astronn\ \citep{Leung2019a} to train a Bayesian convolutional neural network on APOGEE DR14 spectra and asteroseismic ages from APOKASC-2 \citep{Pinsonneault2018}.
Retrained on DR17 spectra, the new catalog improves the performance at low metallicity by incorporating additional asteroseismic data from \citet{Montalban2021} and provides individual stellar abundances \citep{Leung2019a} and distances \citep{Leung2019b} through {\it Gaia}-eDR3 \citep{GaiaCollaboration2021}.
We discuss the differences between \astronn\ and the recent \citet{Leung2023} catalog in Section \ref{sec:empirical:caveats} below.
\par
We filter the sample based on the following selection criteria:
\begin{itemize}

    \itemsep 0pt

    \item \texttt{STAR\_BAD == 0}

    \item \texttt{EXTRATARG == 0}

    \item S/N $\geq 80$

    \item $\log g = 1 - 3.8$

    \item $T_\text{eff} = 3400 - 5500$ K

    \item $\log g < 3$ or $T_\text{eff} > 4000$ K

\end{itemize}
The final criterion excludes the lower-right corner of the Kiel diagram (see Figure \ref{fig:sample} below) to avoid potential contamination by the main sequence.
Age uncertainties in the \astronn\ catalog become substantial for stars older than $\tau \gtrsim 8 - 10$ Gyr \citep{Leung2023}.
Since we are most interested in thin disk populations, we impose additional cuts on age, radius, and midplane distance of:
\begin{itemize}

    \item $\tau \leq 10$ Gyr

    \item $R \leq 15$ kpc

    \item $\left|z\right| \leq 0.5$ kpc,

\end{itemize}
respectively.
These criteria yield a final sample of $N =$ 94,387 red giant and red clump stars.
% Relative to \citet{Lu2022b}, our sample is smaller ($\sim$40\% of its size) but achieves better radial coverage of the disk.
% Dust obscuration at the optical wavelengths of LAMOST limits their sample to the $R \approx 7 - 10$ kpc range.
\par
The left panel of Figure \ref{fig:sample} shows the Kiel diagram of our sample color coded by stellar age.
Young populations are preferentially located in the red clump, whereas the oldest stars distribute themselves more evenly along the red giant branch.
This effect is primarily driven by actual changes in the distribution of stars along the giant branch as a function of population age \citep[e.g.,][]{Girardi2016}.
However, systematic uncertainties in APOGEE abundances tend to present as spurious correlations with $T_\text{eff}$ and $\log g$ \citep[e.g.,][]{Joensson2018, Eilers2022}.
Our main conclusions could be affected by systematics if $T_\text{eff}$ and $\log g$ act as confounding variables in metallicity and age, but this potential issue is not a concern (see discussion in Section \ref{sec:empirical:caveats} below).
% Our main conclusions in this paper are driven by variations in abundances between mono-age populations, which could be affected by systematics if $T_\text{eff}$ or $\log g$ act as confounding variables in metallicity and age (see discussion in Section \ref{sec:empirical:caveats} below).
\par
The right panel of Figure \ref{fig:sample} shows the Galactocentric radii and ages of each star along with the associated 1-D distributions.
Our sample achieves excellent radial coverage of the disk, particularly in the $R = 5 - 12$ kpc range, where 81,580 of the 94,387 stars in our full sample reside.
Coverage is best in the solar annulus, with 30,201 stars found between $R = 7$ and 9 kpc.
Unless otherwise noted, we focus on bins of Galactocentric radius and age that are 1 kpc and 1 Gyr wide throughout this paper.
A consequence of this choice is that neighboring age bins trace truly distinct populations more reliably for young stars than old stars, since the uncertainties tend to be log-normal in shape.
For example, the distributions of the actual ages of stars in our $7 - 8$ and $8 - 9$ Gyr bins likely overlap to a greater extent than our $1 - 2$ and $2 - 3$ Gyr bins.
We therefore color code the right panel of Figure \ref{fig:sample} by the overall age uncertainty $\sigma_\tau$ as opposed to the fractional uncertainty $\sigma_\tau / \tau$ to illustrate the extent to which stars in neighboring bins reliably differ from one another.
% As a consequence of this choice, neighboring age bins trace stellar populations with truly distinct age distributions to a lesser extent for old stars than young stars due to larger uncertainties.
% Stars in our $7 - 8$ and $8 - 9$ Gyr bins, for example, are less different from one another than the $1 - 2$ and $2 - 3$ Gyr bins.
% We therefore color code the right panel of Figure \ref{fig:sample} by the overall age uncertainty $\sigma_\tau$ as opposed to the fractional uncertainty $\sigma_\tau / \tau$ in order to illustrate the extent to which neighboring age bins are reliably distinct from one another.
\par
The reported age uncertainties reach $ \sigma_\tau \approx 1$ Gyr or better across much of the Galactic disk for young populations ($\tau \lesssim 3 - 4$ Gyr) and up to $\tau \approx 10$ Gyr near the Sun.
Stellar abundances have median uncertainties of $\sigma_\onh{Fe} = 0.0087$ and $\sigma_\text{[O/Fe]} = 0.017$, which is sufficiently precise for our purposes.
We discuss sources of uncertainty further in Section \ref{sec:empirical:caveats} below.
% {\color{red}
% (grab this later, it's a good callback but it's redundant here).
% Taken at face value, the reported uncertainties indicate that our measurements most tightly constrain the enrichment history of the disk over the last few Gyr across all radii and over most of the disk lifetime near the sun.
% }

\subsection{Radial Gradients}
\label{sec:empirical:gradients}

The top panels of Figure~\ref{fig:gradients} show age and abundance distributions in our sample in 2-kpc bins of Galactocentric radius.
We have not made any corrections for the survey selection function, so these are distributions of observed APOGEE stars satisfying our selection cuts as opposed to mass-weighted distributions.
In agreement with previous work (see discussion in Section \ref{sec:intro}), stars tend to be young in the outer Galaxy and old in the inner Galaxy, with tails toward old and young populations, respectively.
The metallicity distribution function (MDF) follows a similar pattern, shifting from a metal-rich mode to a metal-poor mode with increasing radius.
The change in the MDF shape, from positively to negatively skewed with radius, was first noted by \citet{Hayden2015} and interpreted as a sign of stellar migration.
The age distribution shown here exhibits a similar change in shape with radius.
\par
We quantify the strength of these radial gradients by computing summary statistics of each distribution in 1-kpc bins of Galactocentric radius.
The bottom left panel of Figure \ref{fig:gradients} shows the median age and the 16th and 84th percentiles of the age distribution as a function of radius.
A linear regression indicates a median trend of $\nabla \tau_{1/2} = -0.375 \pm 0.036$ Gyr/kpc with an intercept of $8.21 \pm 0.31$ Gyr.
The observed median age closely follows this line of best-fit within $R \lesssim 9$ kpc, beyond which ages begin to increase slightly with radius.
\par
While we quantify the radial age gradient in terms of a median trend, we use the mode to quantify the metallicity gradient.
We discuss our motivation behind this choice in Section \ref{sec:discussion:migration} below.
In short, our GCE models suggest that the mode is less susceptible to modification by stellar migration than the mean and median.
As a result, the peak of the MDF is a better proxy for the ISM abundance at a given radius and lookback time.
To mitigate noise in the inferred mode introduced by poisson fluctuations, we first fit a skew normal distribution to the MDF in each radial bin and determine the positions of the peaks with optimization.
\par
The bottom right panel of Figure~\ref{fig:gradients} shows the resultant gradients in~\onh{O}~and~\onh{Fe}.\footnote{
    We follow conventional notation where [X/Y] $\equiv \log_{10} \left(N_\text{X} / N_\text{Y}\right) - \log_{10}\left(N_\text{X} / N_\text{Y}\right)_\odot$.
    % Although gas-phase metallicity measurements are conventionally reported as $12 + \log(N_\text{X} / N_\text{H})$, we transform these values to the same scaling as stars in this paper to better facilitate comparison between the two.
}
As expected, stars tend to decline in metallicity with increasing radius.
Linear regressions indicate slopes of~$\grad{O} = -0.062 \pm 0.001$ kpc$^{-1}$
and~$\grad{Fe} = -0.070 \pm 0.003$ kpc$^{-1}$, in reasonable agreement with
previous measurements from APOGEE \citep[e.g.,][]{Frinchaboy2013, Myers2022}.
For comparison, we additionally plot~\citeauthor{MendezDelgado2022}'s
\citeyearpar{MendezDelgado2022} fit to the gas-phase O gradient traced by
Galactic HII regions including their estimated correction for temperature
inhomogeneities.
The two O gradients are consistent within their~$1\sigma$ uncertainties.
For stars, the [O/H] and [Fe/H] gradients are similar because populations near the midplane typically have [O/Fe] $\approx$ 0.

\subsection{Evolution of the Abundance Gradient}
\label{sec:empirical:evolution}

% fig 3
\begin{figure*}
\centering
\includegraphics[scale = 0.87]{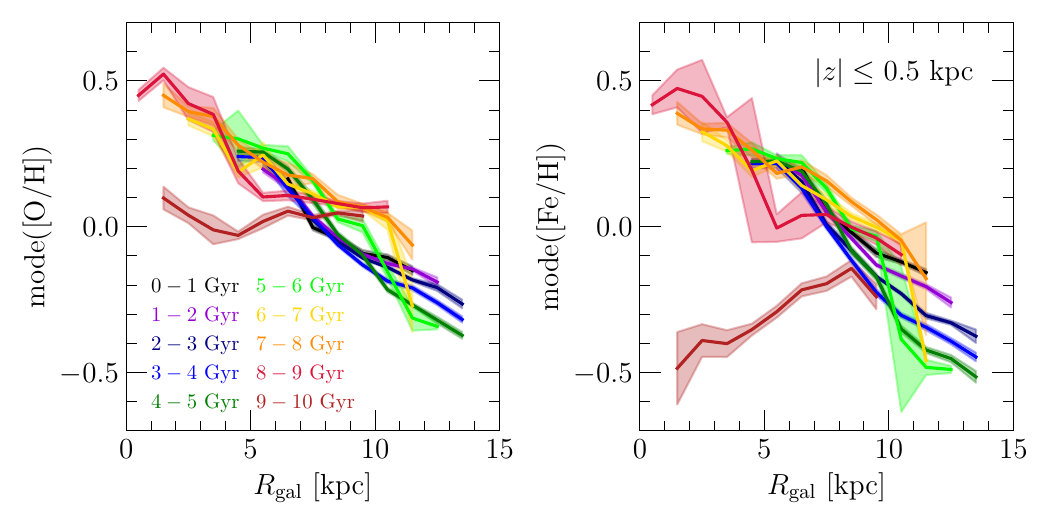}
\includegraphics[scale = 0.82]{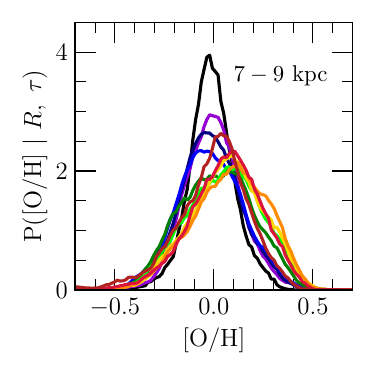}
\includegraphics[scale = 0.82]{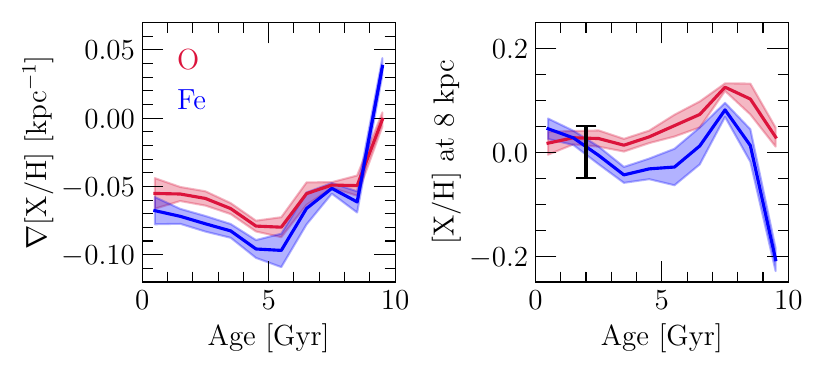}
\caption{
Radial metallicity gradients as a function of stellar age for stars within $\left|z\right| \leq 0.5$ kpc of the disk midplane.
{\bf Top}: The peak of the [O/H] (left) and [Fe/H] (right) distributions in 1-kpc bins of radius, color-coded by stellar age, with young populations in cool colors and old populations in warm colors,  according to the legend in the top-left panel.
Shaded regions denote the statistical uncertainty in the mode of the MDF estimated through jackknife resampling (see discussion in Section \ref{sec:empirical:evolution}).
{\bf Bottom}: The distribution in \onh{O} in the solar annulus ($R = 7 - 9$ kpc; left) in the same age bins as the top panels.
Middle and right panels show the slope and value at $R = 8$ kpc, respectively, inferred from linear regressions to the metallicity gradients in each age bin, parameters of which are reported in Table \ref{tab:regressions}.
The error bar in the right panel marks $0 \pm 0.05$, broadly consistent with \citepossessivepar{Wenger2019} measurement of the non-axisymmetric variations in Galactic HII regions.
{\bf Summary}: In terms of both slope and normalization, variations in the Galactic disk radial metallicity gradient with stellar population age are minimal up to $\sim$9 Gyr, within some dispersion of a time-averaged trend.
% {\bf Summary}: The radial metallicity gradient of stars in the Galactic disk is largely age-independent up to $\sim$$9$ Gyr in terms of both slope and normalization.
}
\label{fig:gradxh-fixedage}
\end{figure*}

\begin{table}
\caption{
A summary of the linear regressions applied to radial metallicity and age gradients in Figures \ref{fig:gradients} and \ref{fig:gradxh-fixedage} (see discussion in Sections \ref{sec:empirical:gradients} and \ref{sec:empirical:evolution}).
We use a pivot point at $R_\text{gal} = 8$ kpc in all regressions (i.e., $y = m(R - R_\odot) + b$).
}
\begin{tabularx}{\columnwidth}{c @{\extracolsep{\fill}} c r}
% \toprule
\hline
Age Range & Slope & Value at $8$ kpc
\\
\hline
\\[-5pt]
\multicolumn{3}{c}{\textbf{[O/H]}}
\\[5pt]
All Stars & $-0.062 \pm 0.001$ kpc$^{-1}$ & $0.028 \pm 0.006$
\\
$\leq 9$ Gyr & $-0.062 \pm 0.002$ kpc$^{-1}$ & $0.028 \pm 0.007$
\\
$0 - 1$ Gyr & $-0.055 \pm 0.011$ kpc$^{-1}$ & $0.018 \pm 0.022$
\\
$1 - 2$ Gyr & $-0.056 \pm 0.005$ kpc$^{-1}$ & $0.028 \pm 0.013$
\\
$2 - 3$ Gyr & $-0.059 \pm 0.005$ kpc$^{-1}$ & $0.026 \pm 0.016$
\\
$3 - 4$ Gyr & $-0.066 \pm 0.004$ kpc$^{-1}$ & $0.014 \pm 0.012$
\\
$4 - 5$ Gyr & $-0.079 \pm 0.004$ kpc$^{-1}$ & $0.030 \pm 0.012$
\\
$5 - 6$ Gyr & $-0.080 \pm 0.007$ kpc$^{-1}$ & $0.051 \pm 0.021$
\\
$6 - 7$ Gyr & $-0.055 \pm 0.008$ kpc$^{-1}$ & $0.073 \pm 0.025$
\\
$7 - 8$ Gyr & $-0.049 \pm 0.002$ kpc$^{-1}$ & $0.125 \pm 0.007$
\\
$8 - 9$ Gyr & $-0.049 \pm 0.007$ kpc$^{-1}$ & $0.103 \pm 0.030$
\\
$9 - 10$ Gyr & $-0.001 \pm 0.005$ kpc$^{-1}$ & $0.029 \pm 0.018$
\\
\hline
\\[-5pt]
\multicolumn{3}{c}{\textbf{[Fe/H]}}
\\[5pt]
All Stars & $-0.070 \pm 0.003$ kpc$^{-1}$ & $-0.019 \pm 0.013$
\\
$\leq 9$ Gyr & $-0.068 \pm 0.003$ kpc$^{-1}$ & $-0.023 \pm 0.014$
\\
$0 - 1$ Gyr & $-0.068 \pm 0.010$ kpc$^{-1}$ & $0.045 \pm 0.020$
\\
$1 - 2$ Gyr & $-0.072 \pm 0.005$ kpc$^{-1}$ & $0.028 \pm 0.014$
\\
$2 - 3$ Gyr & $-0.077 \pm 0.006$ kpc$^{-1}$ & $-0.006 \pm 0.017$
\\
$3 - 4$ Gyr & $-0.083 \pm 0.005$ kpc$^{-1}$ & $-0.044 \pm 0.015$
\\
$4 - 5$ Gyr & $-0.096 \pm 0.007$ kpc$^{-1}$ & $-0.032 \pm 0.020$
\\
$5 - 6$ Gyr & $-0.097 \pm 0.012$ kpc$^{-1}$ & $-0.028 \pm 0.035$
\\
$6 - 7$ Gyr & $-0.066 \pm 0.012$ kpc$^{-1}$ & $0.012 \pm 0.036$
\\
$7 - 8$ Gyr & $-0.051 \pm 0.004$ kpc$^{-1}$ & $0.082 \pm 0.014$
\\
$8 - 9$ Gyr & $-0.061 \pm 0.008$ kpc$^{-1}$ & $0.013 \pm 0.031$
\\
$9 - 10$ Gyr & $\;\;0.038 \pm 0.006$ kpc$^{-1}$ & $-0.207 \pm 0.022$
\\
\hline
\\[-5pt]
\multicolumn{3}{c}{\textbf{Age}}
\\[5pt]
All Stars & $-0.375 \pm 0.036$ Gyr kpc$^{-1}$ & $5.21 \pm 0.16$ Gyr
\\
\hline
\end{tabularx}
\label{tab:regressions}
\end{table}

In this section, we quantify the evolution of the disk abundance structure by repeating our measurements in Section \ref{sec:empirical:gradients} above in 1 Gyr bins of stellar age.
After sorting based on both age and radius, we fit for the mode of the MDF only if the bin contains at least 200 stars.
We estimate statistical uncertainties using jackknife resampling.
\par
The top panels of Figure~\ref{fig:gradxh-fixedage} show the resulting relationship between metallicity and Galactocentric radius for mono-age populations.
While there are some noticeable variations between age bins, the result that is of particular importance to this paper is that the {\it normalization} of this relationship does not obviously change across such a broad range of ages.
To demonstrate this point further, the lower left panel shows the \onh{O} distributions in the solar annulus ($R = 7 - 9$ kpc) in the same 1-Gyr age bins.
There is little to no variation in the peak of the MDF with age.
If anything, the tail of the distribution shifts toward slightly super-solar abundances for old populations, which would only strengthen the argument that old stars are more metal-rich than predicted by conventional GCE models (see discussion in Section \ref{sec:intro}).
\par
We apply linear regressions to these $\onh{O}-R$ and $\onh{Fe}-R$ trends.
The lower middle and lower right panels of Figure \ref{fig:gradxh-fixedage} show the slopes and values at $R = 8$ kpc as functions of the age bin, which we also report in Table \ref{tab:regressions}.
Our measurements indicate that, to first order, both the normalization and slope of this relationship were established $\sim$$8 - 9$ Gyr ago.
For comparison, the error bar in the lower-right panel shows $\onh{O} = 0 \pm 0.05$, which corresponds to the level of non-axisymmetric variations in ISM abundances predicted by simulations \citep{Grand2016} and observed in HII regions \citep{Wenger2019}.
The variations in stellar metallicities up to ages of $\sim$$8 - 10$ Gyr are comparable to this intrinsic dispersion in ISM abundances.
This apparent lack of change in stellar metallicities with age, even for old populations, is our primary motivation in constructing the equilibrium scenario of metallicity gradients.
% As discussed in Section \ref{sec:intro}, similar results have been found using open clusters and classical Cepheids \citep{Spina2022, daSilva2023}, which these measurements extend to significantly higher ages.
\par
Variations in the slope with stellar population age are also small.
Old populations indicate that the gradient first held steady at $\grad{O} \approx \grad{Fe} \approx -0.05$ kpc$^{-1}$ from $\sim$9 to $\sim$6 Gyr ago.
Intermediate-aged populations follow a steeper relation of $\grad{O} \approx -0.07$ kpc$^{-1}$ and $\grad{Fe} \approx -0.09$ kpc$^{-1}$, with the steepest slope occurring at an age of $\sim$5 Gyr.
The gradient then slowly trends back toward its original value of $\grad{O} \approx \grad{Fe} \approx -0.05$ kpc$^{-1}$ toward young ages.
Inspection of the top panels of Figure \ref{fig:gradxh-fixedage} indicates that this steepening is coincident with a decline in stellar metallicities at $R \gtrsim 10$ kpc centered on $\sim$6 Gyr old populations.
In sections \ref{sec:results} and \ref{sec:results:perturbative} below, we argue that these variations are accurately described by an equilibrium state that is perturbed by a merger event.
Increases in the metal abundance with time across all radii predict larger differences between mono-age populations than observed in the MW disk.
Given systematic uncertainties (see discussion in Section \ref{sec:empirical:caveats} below), it is difficult to determine which of the variations between mono-age populations are significant.
The most visually obvious differences, such as these signatures in $\sim$$4 - 6$ Gyr populations, are the most likely to be real.
\par
% In Section \ref{sec:results}, we demonstrate that this nearly age-independent relationship is evidence that the ISM metallicity as a function of radius reached some equilibrium state early in the disk lifetime.
% Small variations in the slope and/or normalization can arise as perturbations thereof (see discussion in Section \ref{sec:results:perturbative}).
% We discuss nuance in the results of Figure \ref{fig:gradxh-fixedage} below.
% Sgr dSph pericentric passage paper: Ruiz-Lara et al. (2020) (already in main.bib file).

\subsection{Sources of Uncertainty}
\label{sec:empirical:caveats}

Because the \astronn\ ages were inferred by modeling the APOGEE spectra (see discussion in Section \ref{sec:empirical:apogee}), the values may be biased by correlations between stellar age and metal abundances (e.g., the age-[O/Fe] relation; \citealt{Feuillet2018, Feuillet2019}).
To assess the impact of this potential systematic uncertainty, we replicate our measurements in Figure \ref{fig:gradxh-fixedage} using the \citet{Leung2023} ages in Appendix \ref{sec:leung2023}.
They demonstrate that their ages are not sensitive to alpha and iron-peak element abundances, making their catalog a useful benchmark for verification.
We find similar results with both catalogs, suggesting that these biases in \astronn\ are not at a level that would alter our main conclusions.
Although the \citet{Leung2023} ages are likely more reliable (see discussion in Appendix \ref{sec:leung2023}), this catalog is a factor of $\sim$$2.5$ smaller due to the narrower range of surface gravities in the training set.
We therefore focus on the \astronn\ ages in this paper, which are reliable in the age range we are most interested in anyway ($\tau \lesssim 8 -  10$ Gyr; see Figure 11 of \citealt{Leung2023} and discussion in their Section 7.1).
\par
Stellar ages themselves, including those that are not based on neural networks, are also a significant source of uncertainty \citep[e.g.,][]{Soderblom2010, Chaplin2013}.
If the uncertainties are particularly large, then the apparent lack of relationship with age seen in Figure \ref{fig:gradxh-fixedage} could arise because different age bins do not actually trace distinct populations.
However, in order for uncertainties to drive this conclusion, the measurements must be sufficiently imprecise to not tell the difference between $\sim$$1 - 3$ and $\sim$$7 - 9$ Gyr old stars.
With or without machine learning, asteroseismology can at least reliably distinguish between young, intermediate, and old populations \citep[e.g.,][]{Leung2023, Stone-Martinez2024}.
The result that metallicity does not decline substantially with age across this broad range should therefore be statistically significant even with the uncertainties involved.
Our confidence in this conclusion is also supported by related investigations using ages estimated with different methods, which have found similar results (\citealt{Spina2022, daSilva2023, Gallart2024}; see discussion in Section \ref{sec:intro}).
\par
Although we have not accounted for target selection in APOGEE (which is described in detail in \citealt{Zasowski2013}, \citeyear{Zasowski2017}, \citealt{Beaton2021}, and \citealt{Santana2021}), these effects also should not affect our main empirical result.
At a given distance, selection effects are introduced into our sample by the metallicity dependence of red giant lifetimes and luminosities, which is minimal at disk-like abundances \citep[e.g.,][]{Hurley2000}.
A representation such as Figure \ref{fig:gradxh-fixedage} should therefore be relatively robust.
In support of this argument, we note that \citet{Imig2023} did account for selection effects.
Their measurements in mono-age populations also do not show an obvious decline in metallicity toward old ages across most of the Galactic disk (see their Figure 17).
\par
Systematic uncertainties in APOGEE abundances may be a concern if $T_\text{eff}$ or $\log g$ act as confounding variables in age or metallicity \citep[e.g.,][]{Joensson2018, Eilers2022}.
The $\log g$ distribution in our sample changes significantly with Galactocentric radius, since low surface gravity stars tend to be more luminous.
This spatial dependence could introduce correlations between metallicity and radius that do not reflect the underlying stellar populations in the thin disk.
These systematics should not affect our main conclusions for two reasons.
First, our conclusions are primarily driven by comparisons between mono-age populations at fixed Galactocentric radius, whose $\log g$ distributions should be similarly affected by distance and selection.
Second, the corrective factors applied to APOGEE abundances to account for these issues are small ($\lesssim$$0.1$ dex; \citealt{Sit2024}).
We also focus on O and Fe throughout this paper, which are not the elements most affected by systematics.

\section{Galactic Chemical Evolution Models}
\label{sec:gce}

{
\renewcommand{\arraystretch}{1.1}
\begin{table*}
\caption{
A summary of our GCE models.
Our naming scheme references the functional form or constant value of the mass loading factor $\eta$ (see Equation \ref{eq:eta-def}) and the overall normalization of stellar yields $y / Z_\odot$ (see Table \ref{tab:yields}).
The primary set (top section; see discussion in Section \ref{sec:gce:primary-set}) differs first and foremost in its treatment of outflows, with $\eta_\odot$ and $R_\eta$ setting the functional form of $\eta$ as a function of radius (see Equation \ref{eq:eta-radius}).
We select \modelname{Exp}{2} as a fiducial equilibrium scenario model, from which we construct informative comparison cases (bottom section; see discussion in Section \ref{sec:gce:variations}).
The $\eta \propto e^R$ models are calibrated to both the observed age and metallicity gradients, while the $\eta = $ constant models are calibrated only to the metallicity gradients (see discussion in Section \ref{sec:gce:calibration}), though the latter approximately reproduce the age gradient anyway (see Figure \ref{fig:calibration-results}).
This difference arises because the $\eta \propto e^R$ models have an additional free parameter ($R_\eta$) that separates the two gradients, while both are determined by the same parameters in the $\eta =$ constant models (i.e., radial variations in the shape of the SFH).
}
\begin{tabularx}{\textwidth}{c @{\extracolsep{\fill}} c c c c c}
\hline
% \hline
{\bf Model Name} & {\bf $\eta_\odot$} & {\bf $R_\eta$} & {\bf $y / Z_\odot$} & {\bf Star formation History} & {\bf Calibrated to Reproduce}
\\
\hline
\hline
\multicolumn{6}{l}{
{\it Primary Set}
}
\\
\hline
\modelname{Exp}{2} & 1.4 & 7 kpc & 2 &
$f_\text{rise-fall}(t)$ &
$\tau_{1/2}(R)$ and [O/H]$_\text{ISM}(R)$
\\
\modelname{Exp}{1} & 0.4 & 7 kpc & 1 & $f_\text{rise-fall}(t)$ & $\tau_{1/2}(R)$ and [O/H]$_\text{ISM}(R)$
\\
% \multicolumn{6}{l}{
% {\it Sub-Equilibrium Models}
% }
% \\
% \hline
\modelname{0.4}{1} & 0.4 & $\infty$ ($\eta = $ constant) & 1 & $f_\text{rise-fall}(t)$ & [O/H]$_\text{ISM}(R)$
\\
\modelname{0}{1} & 0 & $\infty$ ($\eta = $ constant) & 1 & $f_\text{rise-fall}(t)$ & [O/H]$_\text{ISM}(R)$
\\
\multicolumn{6}{l}{
% {\it Slope Variations}
{\it Variations of the \modelname{Exp}{2} model}
}
\\
\hline
\modelname{Exp}{2}-steep & 1.4 & 4.3 kpc ($\nabla_\text{eq} = -0.04$ kpc$^{-1}$) & 2 & $f_\text{rise-fall}(t)$ & N/A
\\
\modelname{Exp}{2}-shallow & 1.4 & 10.9 kpc ($\nabla_\text{eq} = -0.1$ kpc$^{-1}$) & 2 & $f_\text{rise-fall}(t)$ & N/A
\\
% \multicolumn{6}{l}{
% {\it Perturbative \& Sudden Event Models}
% }
% \\
% \hline
% \modelname{evolExp}{3} & 2.4 & {\color{red} $\bigg\lbrace$\parbox{1.2in}{\centering 7 kpc \quad\; ($t \leq 8$ Gyr) \\ 21.7 kpc \: ($t > 8$ Gyr)}} & 3 &
% $f_\text{rise-fall}(t)$ & N/A (built upon \modelname{Exp}{3})
% \modelname{evolExp}{2} & 1.4 & 7 kpc ($t \leq 8$ Gyr), 21.7 kpc ($t > 8$ Gyr) & 2 &
% $f_\text{rise-fall}(t)$ & N/A
% \\
\modelname{Exp}{2}+burst & 1.4 & 7 kpc & 2 & $f_\text{rise-fall}(t)\left(1 + A_b \Phi(t | t_b, \sigma_b, \alpha_b)\right)$ & N/A
\\
\hline
\hline
\end{tabularx}
\label{tab:gcemodels}
\end{table*}
}

{
\renewcommand{\arraystretch}{1.2}
\begin{table}
\caption{A summary of the stellar yields in our GCE models (see Table \ref{tab:gcemodels} and discussion in Section \ref{sec:gce} for details).}
\begin{tabularx}{\columnwidth}{c @{\extracolsep{\fill}} c c}
\hline
Yield & $y / Z_\odot = 1$ & $y / Z_\odot = 2$
\\
\hline
\hline
$\ycc{O}$ & $0.0057$ & $0.0114$
\\
$\yia{O}$ & $0$ & $0$
\\
$\ycc{Fe}$ & $4.5 \times 10^{-4}$ & $9.0 \times 10^{-4}$
\\
$\yia{Fe}$ & $8.4 \times 10^{-4}$ & $16.8 \times 10^{-4}$
\\
\hline
\hline
\end{tabularx}
\label{tab:yields}
\end{table}
}

Our GCE models are adapted from \citetalias{Johnson2021}, which we integrate using the publicly available {\sc Versatile Integrator for Chemical Evolution}\footnote{
    Install: \url{https://pypi.org/project/vice}
    \\
    Documentation: \url{https://vice-astro.readthedocs.io/en/latest}
    \\
    Source code: \url{https://github.com/giganano/VICE.git}
} (\vice; \citealt{Johnson2020}).
Following previous models with similar motivations \citep[e.g.][]{Schoenrich2009a, Minchev2013, Minchev2014}, these models discretize the Galactic disk into $\delta R = 100$ pc annuli from $R = 0$ to $20$ kpc.
By predicting abundances for multiple regions simultaneously, these so-called ``multi-zone'' models are a more realistic description of the Galactic disk than conventional one-zone GCE models (see e.g. the reviews by \citealt{Tinsley1980} and \citealt{Matteucci2021}) and are significantly less computationally expensive than hydrodynamic simulations.
A full detailed description of the framework can be found in section 2 of \citetalias{Johnson2021}, with a shorter summary in section 3 of \citet{Johnson2023a}.
In this section, we review the model components relevant to this paper.
Table \ref{tab:gcemodels} summarizes the key parameters.
\par
These models follow 2,047,550 stellar populations that form over a disk lifetime of $\tau_\text{disk} = 13.2$ Gyr.
We refrain from detailed fits to the disk age-abundance structure observed by APOGEE, instead sticking to a handful of illustrative cases.
We are motivated not by the precision of inferred parameter values but by proof of concept of the equilibrium scenario.
The models we construct are somewhat idealized, treating the assembly history of the MW as one continuous episode of star formation.
This choice allows us to highlight the defining characteristics of equilibrium chemical evolution in disk galaxies.
Simulations predict SFHs to be bursty on $\sim$100 Myr timescales \citep[e.g.,][]{Hopkins2014, Sparre2017, Feldmann2017, Ma2018}, but these features would be washed out on $\sim$Gyr timescales.
This behavior may increase the width of MDFs in mono-age stellar populations (see discussion near the end of Section \ref{sec:discussion:migration:birth-radii}).
Merger events often lead to departures from a smooth SFH sustained on longer timescales, and we investigate one such model in this paper (see discussion in sections \ref{sec:gce:variations} and \ref{sec:results:perturbative}).
\par
Following our measurements in Section \ref{sec:empirical}, we focus on alpha and iron-peak elements, taking O and Fe as representative cases thereof.
Production of these metals is dominated by core collapse supernovae (CCSNe) and Type Ia supernovae (SNe Ia; \citealt{Johnson2019}).
Our yields are defined as the net mass production of either element in units of the mass of the progenitor stellar population.
For example, a value of $y_x = 0.001$ would imply that a hypothetical $1000$ \msun\ star cluster would produce $1$ \msun\ of some element $x$.
In the case of CCSNe, new metals are ejected to the ISM instantaneously, so the rate of change in the surface density of $x$ due to CCSNe follows according to
\begin{equation}
\dot{\Sigma}_x^\text{CC} = \ycc{x} \dot{\Sigma}_\star,
\end{equation}
where $\ycc{x}$ is the population-averaged yield from massive stars and $\dot{\Sigma}_\star$ is the local surface density of star formation.
In the case of SNe Ia, production is spread out over the course of the delay-time distribution (DTD) $R_\text{Ia}$ according to
\begin{equation}
\dot{\Sigma}_x^\text{Ia} = \yia{x} \ddfrac{
    \int_0^t \dot{\Sigma}_\star(t') R_\text{Ia}(t - t') dt'
}{
    \int_0^\infty R_\text{Ia}(t') dt'
},
\end{equation}
where $\yia{x}$ is the population-averaged yield from SNe Ia.
This quantity can be expressed as the product of the mean mass produced by a single SN Ia event and the mean number of events per unit mass of star formation.
We retain the $R_\text{Ia} \propto t^{-1.1}$ single power-law prescription from \citetalias{Johnson2021} based on comparisons between SN Ia rates as a function of redshift and the cosmic SFH \citep[e.g.,][]{Maoz2012a}.
\par
Table \ref{tab:yields} presents our adopted yield values.
With these choices, 35\% (65\%) of Fe arises from CCSNe (SNe Ia) at solar [O/Fe], nearly the same breakdown as \citetalias{Johnson2021}.
This choice places the low [Fe/H] ``plateau'' in [O/Fe] at [O/Fe] $\approx +0.45$, broadly consistent with the abundance distribution observed in APOGEE \citep[e.g.,][]{Hayden2015}.
All of our GCE models follow one of two overall yield normalizations but with the same ratios of yields.
We refer to these two choices collectively as $y / Z_\sun = 1$ and $y / Z_\sun = 2$, because the total yields of O and Fe are either equal to their solar abundances\footnote{
    We use the measurements of O and Fe in the solar photosphere from \citet{Asplund2009}.
} or a factor of two higher.
We show in sections \ref{sec:results} and \ref{sec:discussion:yields:norm} below that the overall scale of stellar yields is related to the timescale on which the disk reaches the equilibrium state.
% {\color{red}
% Try $y / Z_\odot = 2$ and see if this works.
% Original \citetalias{Johnson2021} choice corresponds to $y / Z_\odot = 2.6$.
% Using $y / Z_\odot = 2$ as the fiducial choice would be better if it works (noteworthy if it doesn't) since it is less extreme.
% One advantage of $y / Z_\odot = 3$ though is that the W18F landscape from Emily's black hole landscape paper predicts this level of stellar yields.
% }
\par
These choices of stellar yields are based on a mix of theoretical and empirical considerations.
\citet{Griffith2021b} demonstrate that plausible variations in the amount of black hole formation can account for a factor of $\sim$3 difference in alpha element production.
If most massive stars explode as CCSNe, then the predictions of typical models \citep[e.g.][]{Limongi2018, Sukhbold2016} imply $y / Z_\odot = 2 - 3$.
\citepossessivepar{Rodriguez2023} recent measurement of the mean $^{56}$Ni yield from Type II SNe using the radioactive tails of their lightcurves provides one of the few empirical anchors on the scale of stellar yields.
Their measurements imply $y / Z_\odot \approx 1$ with a modest amount of black hole formation \citep{Weinberg2024}.
We discuss the normalization of stellar yields further in Section \ref{sec:discussion:yields:norm} below.
\par
Following \citetalias{Johnson2021}, our models assume a fiducial choice of SFH given by
\begin{equation}
\dot{\Sigma}_\star \propto f_\text{rise-fall}(t) \equiv
\left(1 - e^{-t / \tau_\text{rise}}\right) e^{-t / \tau_\text{sfh}},
\label{eq:rise-fall-sfh}
\end{equation}
where $\tau_\text{rise}$ and $\tau_\text{sfh}$ control the timescales on which the SFR rises at early times and falls at late times, respectively.
Their values differ between models (see discussion in Section \ref{sec:gce:calibration} and Appendix \ref{sec:calibration}).
This prescription allows more control over the detailed shape of the SFH than a linear-exponential $t e^{-t / \tau_\text{sfh}}$ form at the expense of an additional free parameter.
The normalization of the SFH is set at each radius such that the predicted stellar mass and surface density gradient at the present day are consistent with the findings of \citet{Licquia2015} and \citet{Bland-Hawthorn2016}, respectively.
Combined with our assumed star formation law (see discussion below), this prescription also ensures that the gas surface density profile reasonably resembles that of the MW \citep[e.g.,][]{Kalberla2009}.
In Section \ref{sec:gce:variations} below, we describe a model in which we impose an accretion-induced burst of star formation atop the functional form of Equation \refp{eq:rise-fall-sfh}.
\par
{\sc VICE} computes the accretion rates at each timestep.
The solution is unique given mass conservation and the prescriptions for outflows and star formation efficiency described below.
Mathematically, accretion rates work out to simply make up the difference between what is lost to star formation and outflows and what is required to fuel the specified level of star formation at the next timestep.
We assume accreting material to have zero metallicity, which we relax in Section \ref{sec:discussion:yields:norm} below.
It is inconsequential to instead run {\sc VICE} in ``infall mode,'' directly specifying the accretion history and letting the SFH be computed by the code.
That is, similar predictions arise if we use the accretion histories computed by {\sc VICE} (shown in the bottom row of Figure \ref{fig:allmodels-evol} below) as input to a separate set of models run in infall mode.
\par
Previous iterations of the \citetalias{Johnson2021} GCE models have implemented the radial migration of stars \citep[e.g.][]{Sellwood2002} by ``tagging'' stellar populations with star particles from the {\tt h277} hydrodynamic simulation \citep{Christensen2012} that formed at similar radii and times.
Such an approach can be understood as enforcing the dynamical history of {\tt h277} on the GCE model.
Here, we use the updated version presented in \citeauthor{Dubay2024} (\citeyear{Dubay2024}; see their Appendix C).
Present-day Galactocentric radii are determined by sampling from a normal distribution centered on the birth radius of a stellar population, while mid-plane distances $\left|z\right|$ are determined by sampling from a sech$^2$ function \citep{Spitzer1942}.
This approach closely approximates the distributions of final radius and mid-plane distance seen in {\tt h277} for mono-age populations formed in different Galactic regions.
The advantage over the original tagging approach is that stellar populations born in the outer disk at early times were subject to sampling noise due to the rarity of these populations in the simulation.
\par
We have also updated the star formation law used in these GCE models.
In \citetalias{Johnson2021}, we used a three-component power-law relationship between the surface densities of gas  $\Sigma_g$ and star formation $\dot{\Sigma}_\star$ based on observations by \cite{Bigiel2010} and \citeauthor{Leroy2013} (\citeyear{Leroy2013}; see Figure 2 of \citealt{Krumholz2018b}).
The three component power-law complicates our parameter calibration, which determines $\tau_\text{rise}$ and $\tau_\text{sfh}$ as a function of Galactocentric radius in each of our GCE models (see discussion in Section \ref{sec:gce:calibration} and Appendix \ref{sec:calibration}).
We therefore use a more classical single power-law prescription $\dot{\Sigma}_\star \propto \Sigma_g^N$ with $N = 1.5$ based on \citet{Kennicutt1998} with a transition to a linear star formation law above $10^8\ \msun$ kpc$^{-2}$.
% $\Sigma_g = 10^8\ \msun$ kpc$^{-2}$.
In detail, we implement this relationship by computing the star formation efficiency (SFE) timescale $\tau_\star \equiv \Sigma_g / \dot{\Sigma}_\star$ (referred to as the ``depletion time'' by some authors) according to
\begin{equation}
\tau_\star = \begin{cases}
\tau_\text{mol} & (\Sigma_g \geq 10^8\ \msun\ \text{kpc}^{-2})
\\
\tau_\text{mol} \left(\ddfrac{\Sigma_g}{10^8 \text{kpc}^{-2}}\right)^{-1/2} &
(\Sigma_g < 10^8\ \msun\ \text{kpc}^{-2}),
\end{cases}
\label{eq:tau-star}
\end{equation}
where $\tau_\text{mol}$ is the value of $\tau_\star$ when all of the hydrogen is in the molecular state.
We retain the prescription for $\tau_\text{mol}$ from \citetalias{Johnson2021}, which takes $\tau_\text{mol} = 2$ Gyr at the present day \citep{Leroy2008, Blanc2009} and a $t^{1/2}$ time-dependence based on variations in the $\Sigma_g - \dot{\Sigma}_\star$ relation with redshift \citep{Tacconi2018}.
By imposing a floor at $\tau_\star = \tau_\text{mol}$, our models do not allow the ISM to form stars more efficiently than observed in molecular gas.
\par
We choose this threshold of $\Sigma_g = 10^8\ \msun$ kpc$^{-2}$ because our model reaches these surface densities only in the inner $\sim$1 kpc where the central molecular zone is found \citep{Morris1996, Dahmen1998, PiercePrice2000, Hatchfield2020} and therefore where $\tau_\star = \tau_\text{mol}$ would most plausibly occur.
As a consequence of this recipe, our models generally have less efficient star formation than \citetalias{Johnson2021} ($\tau_\star \approx 6$ Gyr versus $\sim$3 Gyr at $R = 8$ kpc at the present day).
Observationally, the molecular fraction, $\text{H}_2 / (\text{H}_2 + \text{HI})$, reaches unity near $10^6\ \msun$ kpc$^{-2}$ \citep{Bigiel2008, Blanc2009}, which is two orders of magnitude lower than our prescription.
However, such models predict the ISM to be in the molecular phase across most of the Galactic disk.
Empirically calibrated star formation laws based on population-averaged trends led to similarly large molecular gas fractions in \citetalias{Johnson2021} (see Figure 5 and discussion in Section 2.6 therein), potentially in tension with the observed presence of HI as close to the Galactic center as $\sim$500 pc \citep[e.g.,][]{Kalberla2009}.
Modifications to this prescription, however, do not significantly alter our model predictions, so we do not pursue this question further.
We elect to present models placing the transition at $10^8\ \msun$ kpc$^{-2}$, since this choice leads to variations in SFE across the range of Galactocentric radius where most of our sample is found (see discussion in Section \ref{sec:empirical:apogee}).

\subsection{The Primary Set}
\label{sec:gce:primary-set}

Our ``primary set'' of GCE models is so-named because they highlight the difference between equilibrium and evolution scenarios.
In the evolution scenario, metallicity grows with time until the present day as opposed to reaching some steady state early in the disk lifetime.
However, the equilibrium metallicity is still well defined even in the evolution scenario (see Equation \ref{eq:zoeq-waf17} below).
The key difference between the two is not in regards to the existence of an equilibrium state, but whether or not it is reached within the disk lifetime.
\par
Following models of the MZR in the literature (e.g. \citealt{Finlator2008, Peeples2011}; see discussion in Section \ref{sec:intro}), the models in our primary set differ first and foremost in their prescription for outflows from the Galactic disk.
% If this description is accurate, then the mass loading factor $\eta \equiv \dot{\Sigma}_\text{out} / \dot{\Sigma}_\star$ relating the outflow and star formation surface densities should scale with stellar mass approximately as $\eta \propto M_\star^{-1/3}$ \citep[e.g.][]{Muratov2015, Chartab2023}.
\citet{Weinberg2017b} demonstrate that the equilibrium abundance of O in a one-zone GCE model is given by
\begin{equation}
Z_\text{O,eq} = \frac{\ycc{O}}{1 + \eta - r - \tau_\star / \tau_\text{sfh}},
\label{eq:zoeq-waf17}
\end{equation}
where $\eta$ is the mass loading factor relating the outflow and star formation surface densities:
\begin{equation}
\eta \equiv \frac{
    \dot{\Sigma}_\text{out}
}{
    \dot{\Sigma}_\star
}.
\label{eq:eta-def}
\end{equation}
The corrective term $r$ accounts for the return of stellar envelopes back to the ISM, which can be approximated as a constant in most cases ($r \approx +0.4$ for a \citealt{Kroupa2001} IMF).
If the rates of accretion and ejection are significant compared to star formation, then the factor of $\tau_\star / \tau_\text{sfh}$ is a small correction, and the value of $Z_\text{O,eq}$ is determined by stellar yields and the outflow mass loading $\eta$.
In the limit of zero accretion, Equation \refp{eq:zoeq-waf17} is inaccurate.
$Z_\text{O,eq}$ instead diverges because all of the H is eventually processed by stars and converted into metals (similar to classical ``closed box'' models of GCE; see the review by \citealt{Tinsley1980}).
% $Z_\text{O,eq}$ diverges in the limit of zero accretion (similar to classical ``closed box'' models of GCE; see the review by \citealt{Tinsley1980}), since all of the H is eventually processed by stars and converted into metals.
\par
Equation \refp{eq:zoeq-waf17} indicates that if $\eta$ scales exponentially with radius, then the equilibrium abundance will decline roughly (though not exactly) exponentially with radius, tracking the observed metallicity gradient shape \citep[e.g.][]{Wyse1989, Zaritsky1992}.
We therefore adopt the following parameterization in two of our primary set models:
\begin{equation}
\eta = \eta_\odot e^{(R - R_\odot) / R_\eta},
\label{eq:eta-radius}
\end{equation}
where $\eta_\odot$ sets the value of $\eta$ at the solar radius ($R_\odot = 8$ kpc).
If $\eta$ dominates denominator of Equation \refp{eq:zoeq-waf17}, then the scale radius $R_\eta$ sets the slope of the equilibrium gradient according to
\begin{equation}
\grad{O}_\text{eq} \approx \frac{-1}{%
R_\eta \ln 10
}.
\end{equation}
We use $R_\eta = 7$ kpc as a fiducial value.
Under the above approximation, this choice corresponds to a slope consistent with our measurement of $\grad{O} = -0.062 \pm 0.001$ kpc$^{-1}$.
However, the quantity $1 - r - \tau_\star / \tau_\text{sfh}$ also influences the equilibrium abundance as a function of radius.
We discuss the consequences of our $\eta \propto e^R$ formalism below.
Although $\eta$ is highest in the outer disk, these models still predict the surface density of the outflowing material to be highest at $R = 0$ because $\dot{\Sigma}_\star$ declines with radius faster than $\eta$ increases.
\par
This paper presents two models with the $\eta \propto e^R$ scaling, which we refer to as \modelname{Exp}{1} and \modelname{Exp}{2} based on their assumptions regarding $\eta$ and the scale of stellar yields.
The former assumes the $y / Z_\odot = 1$ normalization, while the latter uses the $y / Z_\odot = 2$ scale.
These models also use a different normalization of the exponential scaling of $\eta$ with radius, namely $\eta_\odot = 0.4$ and $\eta_\odot = 1.4$, but both use the fiducial scale radius of $R_\eta = 7$ kpc.
This additional parameter difference is necessary, because the overall scale of stellar yields is strongly degenerate with the strength of mass loading in GCE models \citep[e.g.][]{Cooke2022, Johnson2023b, Sandford2024}.
These choices of $\eta_\odot$ imply $Z_\text{O,eq} = Z_{\text{O},\odot}$ at $R = 8$ kpc in the limit of a constant SFH (i.e. $\tau_\text{sfh} \rightarrow \infty$; see Equation \ref{eq:zoeq-waf17}).
\par
As comparison cases in the evolution scenario, we also follow models with $\eta = 0$ and $\eta = 0.4$ at all radii (i.e. the limit that $R_\eta \rightarrow \infty$).
Both of these models assume the $y / Z_\odot = 1$ scale of stellar yields.
Following the same naming scheme, we refer to these models as \modelname{0}{1} and \modelname{0.4}{1}.
We highlight the $\eta = 0$ scenario as a particularly interesting comparison, because many GCE models of the MW assume no outflows \citep[e.g.][]{Minchev2013, Minchev2014, Spitoni2019, Spitoni2021, Palla2020, Palla2022, Palla2024, Gjergo2023}.
We discuss mass loading in GCE models of the MW further in Section \ref{sec:discussion:outflows} below.
\par
We demonstrate in Section \ref{sec:results} below that the \modelname{Exp}{2} model is a prototypical example of equilibrium chemical evolution because it predicts ISM metallicities to reach a steady state nearly 10 Gyr ago.
This prediction makes this model useful for highlighting the defining features of the equilibrium scenario.
We therefore select \modelname{Exp}{2} as our fiducial set of parameters.
We define a few simple variations of this base model in Section \ref{sec:gce:variations} below.
\par
Our simple exponential scaling of $\eta$ as a function of radius described above leads to some minor discrepancies with our sample.
Both the \modelname{Exp}{1} and \modelname{Exp}{2} models underpredict abundances by $\sim$$0.1 - 0.2$ dex in the gas-phase across the Galactic disk and in stars at $R \lesssim 7$ kpc (see Figures \ref{fig:calibration-results} and \ref{fig:totalgradxh-datacomp}).
The difference at small radii is caused by a slight flattening of the equilibrium gradient toward the Galactic center, where other terms in the denominator of Equation \refp{eq:zoeq-waf17} are more significant.
The underprediction of the ISM abundance across the disk arises becaues the local equilibrium metallicity is not perfectly static and instead slowly increases due to star formation becoming less efficient (i.e. $\tau_\star$ increasing; see discussion following Equation \ref{eq:zoeq-waf17}), but the evolution is small over the disk lifetime.
Neither of these shortcomings affect our main conclusions.
The key prediction by our fiducial model is that it best explains the lack of evolution in the stellar metallicity gradient with age.

\subsection{Parameter Calibration for the Primary Set}
\label{sec:gce:calibration}

\begin{figure*}
\centering
\includegraphics[scale = 0.98]{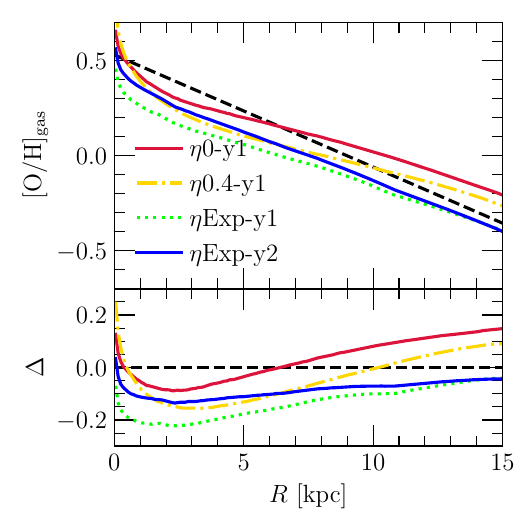}
\includegraphics[scale = 0.99]{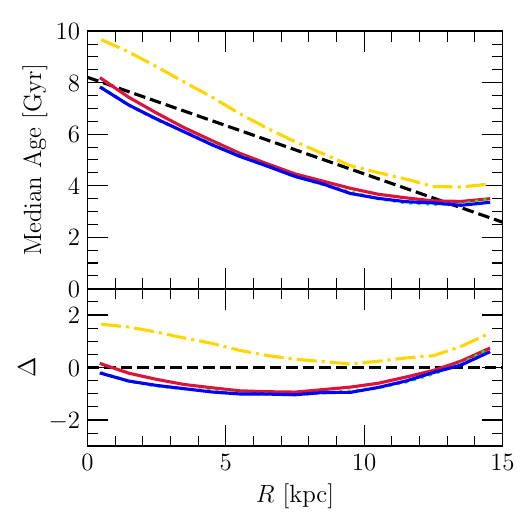}
\caption{
Calibration of our primary set of GCE models (see Table \ref{tab:gcemodels} as well as discussion in Section \ref{sec:gce} and Appendix \ref{sec:calibration}).
{\bf Left}: The predicted [O/H] gradients in the ISM at the present day (colored and styled lines marked by the legend).
The black dashed line marks \citepossessivepar{MendezDelgado2022} measurements in HII regions.
The bottom panel shows the differences between the predicted and observed abundances as a function of radius.
{\bf Right}: The same as the left panel, but for the stellar age gradient.
The black dashed line marks our linear regression in the bottom left panel of Figure \ref{fig:gradients} (see fit parameters in Table \ref{tab:regressions}).
{\bf Summary}: All models are calibrated to reproduce the observed ISM abundances at the present day, which they achieve sufficiently accurately for our purposes.
Only the models invoking $\eta \propto e^R$ are calibrated to reproduce the observed age gradient, but the other models still reasonably reproduce it.
}
\label{fig:calibration-results}
\end{figure*}

Each model in our primary set closely approximates the present-day O abundance in the ISM by construction.
In the \modelname{0}{1} and \modelname{0.4}{1} models, we tune the parameters $\tau_\text{rise}$ and $\tau_\text{sfh}$ describing the shape of the SFH (see Equation \ref{eq:rise-fall-sfh}) such that this observational result is reproduced.
With other parameters held fixed, longer values of $\tau_\text{rise}$ or $\tau_\text{sfh}$ lead to lower [O/H] at $t = \tau_\text{disk} = 13.2$ Gyr.
We describe this procedure in detail in Appendix \ref{sec:calibration}.
In short, we first assume $\tau_\text{rise} = 2$ Gyr as in \citetalias{Johnson2021} and search for a value of $\tau_\text{sfh}$ between $100$ Myr and $200$ Gyr that predicts the observed ISM abundance in an analytic one-zone GCE model.
If no solution is found, we hold $\tau_\text{sfh}$ fixed at $200$ Gyr and search for a value of $\tau_\text{rise}$ between $2$ Gyr and $2 \tau_\text{disk} = 26.4$ Gyr.
If still no solution is found, we simply use the values $\tau_\text{sfh} = 200$ Gyr and $\tau_\text{rise} = 26.4$ Gyr, which corresponds to an SFH that rises approximately linearly with time up to the present day.
We apply this parameter fitting procedure in each radial zone.
% In practice, we find that simply letting the calibration fail and adopting these values directly is accurate enough for our purposes.
% The predicted ISM abundances do not differ from the observed values to an extent that would affect our conclusions (see Figure \ref{fig:calibration-results} and discussion below).
\par
In the \modelname{Exp}{1} and \modelname{Exp}{2} models, the present-day ISM abundances are much more sensitive to the exponential scaling of $\eta$ with radius than they are to the shape of the SFH (see discussion in Section \ref{sec:gce:primary-set} above).
These models therefore have more freedom than the $\eta = $ constant models in this regard, but at the expense of an additional free parameter.
We therefore assign $\tau_\text{rise}$ and $\tau_\text{sfh}$ in the \modelname{Exp}{1} and \modelname{Exp}{2} models such that they reproduce the observed median stellar age as a function of Galactocentric radius.
We describe this procedure in detail in Appendix \ref{sec:calibration} as well.
A given combination of $\tau_\text{rise}$ and $\tau_\text{sfh}$ is accepted if the integral of the implied SFH up to the {\it observed} median age is equal to half of the integral up to the disk lifetime.
Otherwise, we follow the same strategy as with the \modelname{0}{1} and \modelname{0.4}{1} models, starting with $\tau_\text{rise} = 2$ Gyr and searching for a value of $\tau_\text{sfh}$ between $100$ Myr and $200$ Gyr.
Although the observed age gradient is likely affected by selection effects in APOGEE, which we do not account for in this paper (see discussion in Section \ref{sec:empirical:caveats}), this uncertainty is not a major concern for our investigation.
We aim to construct models that predict a plausible MW-like stellar disk but not necessarily a detailed match to the data.
\par
Figure \ref{fig:calibration-results} shows the results of this parameter calibration: the predicted ISM metallicities (left) and median stellar ages (right) as functions of Galactocentric radius for each model in our primary set.
As intended, each model approximately reproduces the observed metal abundances in the ISM.
The $\eta \propto e^R$ models undershoot the observed gradient slightly because the quantity $1 - r - \tau_\star / \tau_\text{sfh}$ also influences the local equilibrium abundance (see Equation \ref{eq:zoeq-waf17}).
As a consequence, the equilibrium state predicted by these models is slightly more metal-poor than present-day ISM abundances, but this discrepancy does not affect our main conclusions (see discussion in Section \ref{sec:gce:primary-set} above).
Since the shape of the SFH is tuned to reproduce the observed stellar age in the \modelname{Exp}{1} and \modelname{Exp}{2} models, any deviations should be a consequence of radial migration.
In our case, the relation was altered from a purely linear trend to one that is slightly concave up.
The observed age gradient is also reasonably approximated by the \modelname{0}{1} and \modelname{0.4}{1} models, even though they were not explicitly calibrated to agree with this trend.
\par
In our equilibrium scenario models, our parameter calibration procedure finds a solution for neither $\tau_\text{rise}$ nor $\tau_\text{sfh}$ at $R \gtrsim 10.5$ kpc (see Figure \ref{fig:calibration} in Appendix \ref{sec:calibration}).
For these cases, we simply adopt $(\tau_\text{rise}, \tau_\text{sfh}) = (26.4, 200)$ Gyr.
This failure arises because the observed median age is low in the outer disk, and it is challenging to parameterize a single-epoch SFH using Equation \refp{eq:rise-fall-sfh} that begins $\tau_\text{disk} = 13.2$ Gyr ago but forms most of its stars at late times.
This discrepancy between these models and the data is not a concern because we are much more interested in their predicted metallicity gradients, which are much more sensitive to the mass loading factor $\eta$.
Our parameter calibration does not face such issues with the evolution scenario models.
The \modelname{0.4}{1} model finds a solution at all radii, and the \modelname{0}{1} model resorts to adopting $(\tau_\text{rise}, \tau_\text{sfh}) = (26.4, 200)$ Gyr only at $R \gtrsim 13.5$ kpc.
% In the $\eta = $ constant models, our calibration procedure does not find a solution for $\tau_\text{rise}$ and $\tau_\text{sfh}$ at $R \gtrsim 7$ kpc and $R \gtrsim 10$ kpc, respectively.
% For these cases, we simply let the calibration procedure fail and adopt the values of $\tau_\text{sfh} = 200$ Gyr and $\tau_\text{rise} = 2$ Gyr.
% This failure arises because it is challenging to find parameters that keep the metallicity in the ISM as low as observed in the outer disk after 13.2 Gyr of star formation.
% As a result, these models overpredict the observed ISM abundance in these regions by a small amount, shown by the residuals in the bottom panels.
% In practice, these parameter choices are accurate enough for our purposes anyway, and calibration difficulties are not the reason for any of the successes or failures of these models.
% In our $\eta \propto e^R$ models, however, there are no radii at which the parameter calibration fails.

\subsection{Variations of the Fiducial Equilibrium Model}
\label{sec:gce:variations}

\begin{figure*}
\centering
\includegraphics[scale = 0.8]{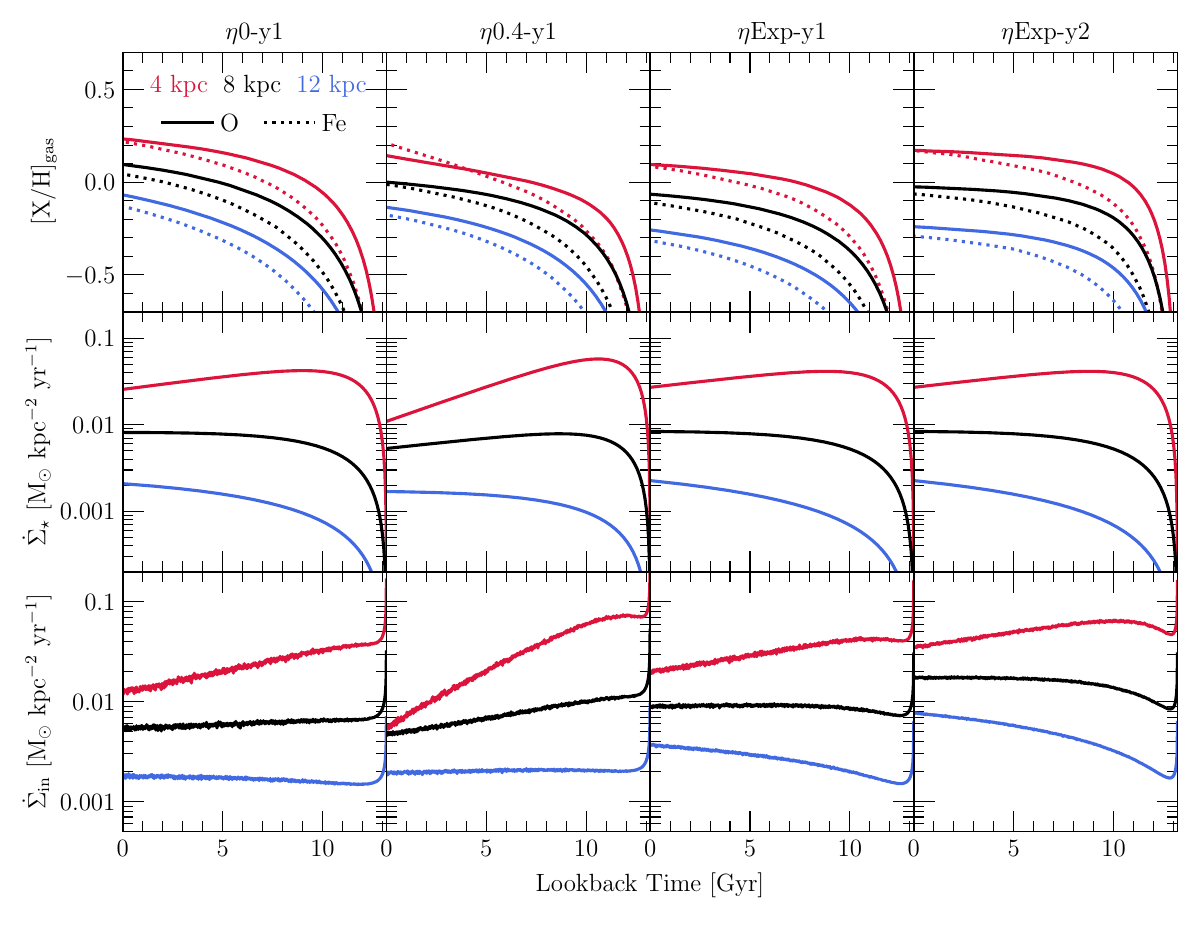}
\caption{
Enrichment (top), star formation (middle), and infall histories (bottom) for our primary set of GCE models, labeled at the top of each column of panels (see Table \ref{tab:gcemodels} and discussion in Section \ref{sec:gce}).
We visualize the evolution at Galactocentric radii of $R = 4$ (red), 8 (black), and 12 kpc (blue) in all panels.
O and Fe abundances in the top row are distinguished with solid and dotted lines, respectively.
{\bf Summary}: Each model approximately reproduces the radial metallicity gradient observed in HII regions (\citealt{MendezDelgado2022}; see Figure \ref{fig:calibration-results} in Appendix \ref{sec:calibration}), but evolves to that point differently due to differences in GCE parameters.
% {\color{red} Try putting a vertical grey line at $\tau = 9$ Gyr and note it as the time our measurements indicate the disk reached equilibrium.}
}
\label{fig:allmodels-evol}
\end{figure*}

% {\color{red}
% Briefly call out slow and fast versions of the radial migration speed at the end of this section.
% Discussion of the \modelname{evolExp}{3} model should obviously be removed if that model gets cut.
% I'm inclined to cut it, because it doesn't seem to get much discussion with the updates to the text, and the main point can be stated in words relatively concisely.
% Also state in words that we follow variations with 25\% faster and 25\% slower radial migration in Section \ref{sec:discussion:migration} and Appendix \ref{sec:birth-radii}.
% Also call out the metal-rich accretion and metallicity dependent yield models in Section \ref{sec:discussion:yields-outflows}.
% }

We show below that ISM abundances in the \modelname{Exp}{2} model rise quickly to equilibrium values, while those in the other models climb steadily over time.
The \modelname{Exp}{2} model therefore achieves the best agreement with the slow evolution shown in Figure \ref{fig:gradxh-fixedage}, so we adopt it as our fiducial model.
We examine several variations of this model in addition to the others in our primary set discussed above.
\par
These variants start from the \modelname{Exp}{2} model parameters and lead to informative comparisons.
Two of these variations have different equilibrium slopes, $\grad{O}_\text{eq}$, which we achieve by simply adjusting the value of $R_\eta$ (see Equation \ref{eq:eta-radius} and discussion in Section \ref{sec:gce:primary-set} above).
The \modelname{Exp}{2}-steep model has a lower value of $R_\eta = 4.3$ kpc, which corresponds to a steeper gradient of $\grad{O}_\text{eq} = -0.1$ kpc$^{-1}$.
The \modelname{Exp}{2}-shallow is the opposite case, with a higher value $R_\eta = 10.9$ kpc and a shallower slope of $\grad{O}_\text{eq} = -0.04$ kpc$^{-1}$.
In Section \ref{sec:discussion:migration} below, we also present models in which we adjust the speed of radial migration.
\par
% Our third variations a model in which $R_\eta$ suddenly changes at time $t = 8$ Gyr ($\tau = 5.2$ Gyr ago) from its initial value of $7$ kpc to $21.7$ kpc (i.e. $\grad{O}_\text{eq} = -0.02$ kpc$^{-1}$).
% This \modelname{evolExp}{3} model describes a scenario in which some property of the Galaxy changes in a manner that affects the long-term equilibrium state.
% We construct this model with no knowledge or assumption about what may cause such a change in nature.
% Our motivation is to predict what may happen if some intrinsic property of the Galaxy changes in a way that affects GCE parameters.
% \par
Our final comparison case is the \modelname{Exp}{2}+burst model, which imposes a burst of star formation in the outer disk $\sim$6 Gyr ago.
The SFH in this model is given by
\begin{equation}
\dot{\Sigma}_\star \propto f_\text{rise-fall}(t) \Big[
1 + A_b \Phi(t | t_b, \sigma_b, \alpha_b)
\Big],
\label{eq:burst-sfh}
\end{equation}
where $\Phi$ is a skew-normal function of time $t$ centered on $t_b$ with standard deviation $\sigma_b$ and skewness $\alpha_b$, and $A_b$ is a dimensionless parameter describing the strength of the burst.
The model we present uses $t_b = 7$ Gyr, $\sigma_b = 1$ Gyr, and $\alpha_b = 3$, with $A_b$ given by
\begin{equation}
A_b = \begin{cases}
0 & (R \leq 4 \text{ kpc})
\\
\min \left[2, e^{(R - 4 \text{ kpc}) / 5 \text{ kpc}} - 1\right] & (R > 4 \text{ kpc}).
\end{cases}
\end{equation}
The burst starts small in amplitude at $R = 4$ kpc and grows in strength with radius, stopping at a factor of 3 increase in the SFR at $R \gtrsim 9$ kpc.
These choices are {\it ad hoc}; we isolated the values by trial and error, choosing them because they approximately reproduce the scale and timing of the variations in abundances seen in intermediate aged stars at $R \gtrsim 10$ kpc in Figure \ref{fig:gradxh-fixedage}.
We discuss this model along with potential origins of a burst of star formation in the outer disk in Section \ref{sec:results:perturbative} below.

\section{Results}
\label{sec:results}

\begin{figure*}
\centering
\includegraphics[scale = 0.9]{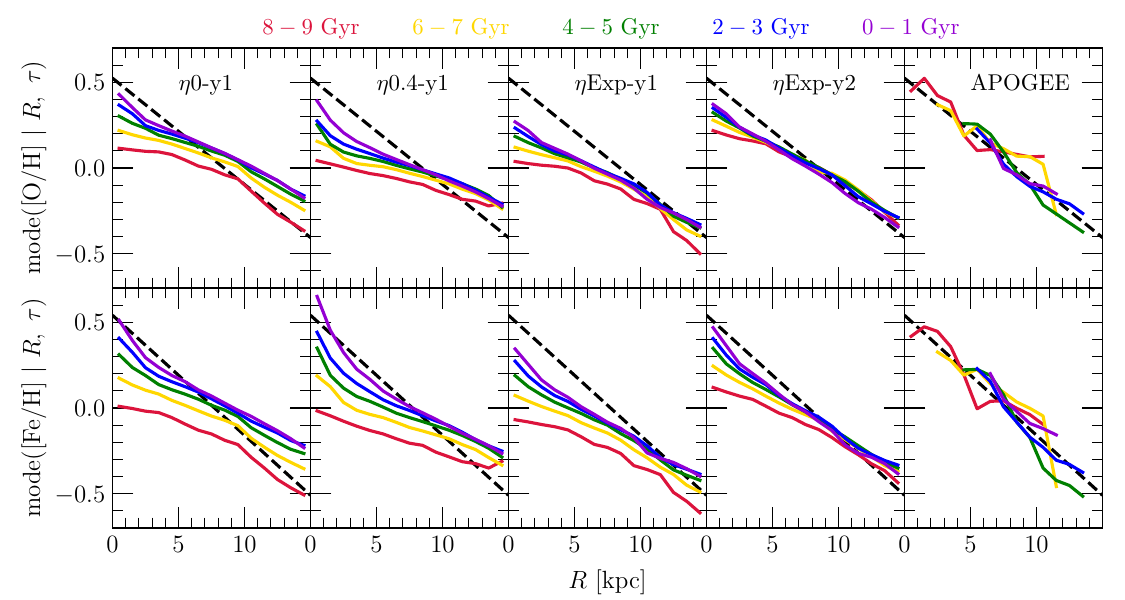}
\caption{
Comparing metallicity gradients in O (top) and Fe (bottom) for different ages of stellar populations (color coded according to the legend at the top).
The first four columns of panels from the left show the predictions of our primary set of GCE models (see discussion in Section \ref{sec:gce:primary-set}), labeled along the top row.
The right hand panels shows our measurements with the \astronn\ value added catalog (see discussion in Section \ref{sec:empirical:evolution}) and are identical to the corresponding points in Figure \ref{fig:gradxh-fixedage}.
The black dashed lines show our measurements of the gradients of all stars, identical to the red and blue curves in the bottom right panel of Figure \ref{fig:gradients}, and are the same across the top and bottom rows of panels.
% Comparing metallicity gradients in O (top) and Fe (bottom) for different ages of stellar populations (color coded according to the legend at the top) between our measurements and our models.
% Stars mark our measurements with the \astronn\ value added catalog (see discussion in Section \ref{sec:empirical}) and are identical to the corresponding points in Figure \ref{fig:gradxh-fixedage}.
% Lines mark the corresponding measurements from our primary set of GCE models, denoted by the text in the top row of panels (see Table \ref{tab:gcemodels} and discussion in Section \ref{sec:gce}).
{\bf Summary}: While agreement is not perfect, the \modelname{Exp}{2} model best reproduces the observed lack of evolution in the metallicity-radius relation with stellar age.
}
\label{fig:allmodels-gradxh-agebins}
\end{figure*}

Figure \ref{fig:allmodels-evol} shows the enrichment, star formation, and accretion histories predicted by our primary set of models (see Table \ref{tab:gcemodels} and discussion in Section \ref{sec:gce:primary-set}).
% These models differ primarily in their treatment of outflows from the Galactic disk.
Each model reaches similar metal abundances in the ISM at the present day (see discussion in Section \ref{sec:gce:calibration} and Appendix \ref{sec:calibration}).
What sets the models apart is that they reach these abundances on different timescales, which leaves observable signatures in the metallicities of mono-age populations.
\par
In the \modelname{0}{1} model, the late-time SFH is gently declining in the inner disk, nearly flat at $R = 8$ kpc, and slightly rising at large radii.
A declining SFH leads to higher metallicity because stellar enrichment is deposited in a declining gas supply.
The trend of SFH shape with radius is stronger in the \modelname{0.4}{1} model so that this gas supply effect can compensate for the loss of metals in outflows.
The \modelname{Exp}{1} and \modelname{Exp}{2} models have the same SFH by construction, and these histories are similar to the \modelname{0}{1} model.
However, the gas accretion rates are higher in the \modelname{Exp}{2} model because the outflows are stronger ($\eta_\odot = 1.4$ versus $0.4$), so more accretion is required to maintain the gas supply required to fuel the stellar mass budget.
\par
% The SFH in the outer disk is quite extended in the \modelname{0}{1} and \modelname{0.4}{1} models.
% This prediction arises due to our parameter calibration, which determines $\tau_\text{rise}$ and $\tau_\text{sfh}$ (see Equation \refp{eq:rise-fall-sfh}) such that the present-day ISM metallicities are closely approximated.
% A steadily increasing accretion rate keeps the metallicities low by ensuring a substantial supply of fresh hydrogen.
% Even though the \modelname{Exp}{1} and \modelname{Exp}{2} models have the same SFH, the accretion history is a factor of $\sim$$3$ higher in the latter.
% This difference is a direct consequence of the higher mass loading factor in the \modelname{Exp}{2} model.
% With stronger outflows from the ISM, more accretion is required to fuel a prescribed level of star formation and fulfill the stellar mass budget (see discussion in Section \ref{sec:results:sfr-per-ifr} below).
% \par
To test these models against the observed metallicity-radius relation in mono-age populations, we apply the same procedure described in Section \ref{sec:empirical:evolution} to our model predictions.
The only difference is that we must weight the metallicity distributions in each bin of radius and age by the mass of each stellar population before computing the position of the mode.
This additional step is necessary, because the mass of each stellar population determines the number of stars that would be available in an APOGEE-like survey.
\par
Figure \ref{fig:allmodels-gradxh-agebins} shows the results of this procedure for both O and Fe.
The \modelname{Exp}{2} model best reproduces the empirical result that populations of different ages follow a similar relation between metallicity and Galactocentric radius.
The agreement is not perfect; most notably, this model predicts some subtle differences in \onh{Fe} between age bins at $R \lesssim 5$ kpc, but it performs sufficiently well for a qualitative reproduction.
The \modelname{0.4}{1} and \modelname{Exp}{1} models perform reasonably well at $R \gtrsim 10$ kpc, but at smaller radii, there are significant variations between age bins that are not present in the observations.
The \modelname{0}{1} model performs the worst in this comparison.
While this model has poor agreement with the observed gradient slope, we consider its more serious shortcoming to be predicting a global increase in the normalization of the metallicity-radius relation by $\sim$$0.3 - 0.4$ dex over this age range, contrary to the observations.
% \par
The model-to-model differences in Figure \ref{fig:allmodels-gradxh-agebins} follow the expectations from the enrichment histories in Figure \ref{fig:allmodels-evol}, but Figure \ref{fig:allmodels-gradxh-agebins} includes the impact of radial migration on the stellar metallicity gradients.
% For all models, the predicted evolution of Fe gradients is stronger than that of O gradients because SN Ia time delays slow the approach to equilibrium (dotted curves in the top row of Figure \ref{fig:allmodels-evol}).
% For the same reason, our \modelname{Exp}{2} struggles to reach [Fe/H] $\approx 0.3 - 0.5$ in the inner disk as early as $\sim$9 Gyr ago as suggested by our sample.
% Non-zero metallicity accretion and/or metallicity dependent SN rates may help increase [Fe/H] in old populations in future versions of these models (see also discussion in Section \ref{sec:discussion:yields:norm}).

\begin{figure*}
\centering
\includegraphics[scale = 0.85]{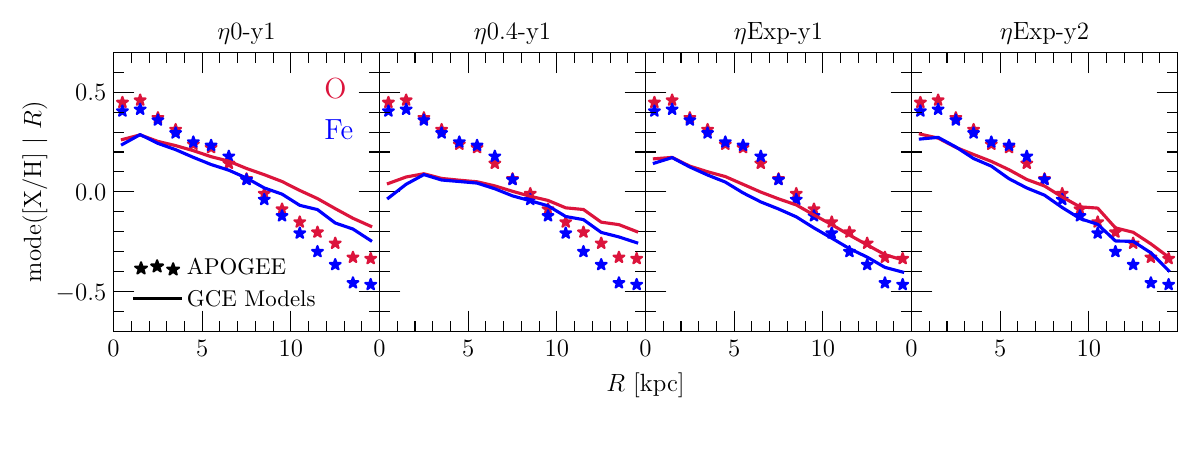}
\caption{
Comparing predicted stellar metallicity gradients with our empirical measurements.
Star symbols denote our measurements from APOGEE in O (red) and Fe (blue) and are identical to the points in Figure \ref{fig:gradients} (see discussion in Section \ref{sec:empirical:evolution}).
Lines denote the predictions from our primary set of GCE models, marked at the top of each panel (see Table \ref{tab:gcemodels} and discussion in Section \ref{sec:gce}).
{\bf Summary}: Since the models are calibrated to reproduce the observed gradient in the ISM, it is not guaranteed that they will also reproduce the observed stellar gradient.
The \modelname{Exp}{2} model offers the best explanation of the consistency between gradients in gas and stars seen in Figure \ref{fig:gradients}.
}
\label{fig:totalgradxh-datacomp}
\end{figure*}

In Figure \ref{fig:totalgradxh-datacomp}, we compare the predicted metallicity-radius relation for all stars (i.e., without sorting into age bins) with our measurements from the lower-right panel of Figure \ref{fig:gradients}.
Because our models are calibrated to reproduce the present-day ISM abundances, it is not guaranteed that they will also reproduce the stellar gradient.
This comparison therefore isolates the model that best reproduces the observed consistency between the metallicities of gas and stars as a function of radius (see discussion in Section \ref{sec:empirical:gradients}).
Once again, the \modelname{Exp}{2} model performs the best.
The \modelname{0}{1} model is a close second, with a slope that is only slightly shallower than that of \modelname{Exp}{2}.
The \modelname{Exp}{1} and \modelname{0.4}{1} models fail most noticeably in the inner disk, where they underpredict stellar abundances significantly more than the \modelname{Exp}{2} and \modelname{0}{1} models.
\par
Holistically, our \modelname{Exp}{2} model offers the best agreement with the data when Figures \ref{fig:allmodels-gradxh-agebins} and \ref{fig:totalgradxh-datacomp} are both considered, but agreement is not perfect.
This model underpredicts stellar metallicities at $R \lesssim 7$ kpc by $\sim$$0.1$ dex, largely due to subtle differences in shape between the equilibrium gradient and the observed gradient (see discussion at the end of Section \ref{sec:gce:primary-set}).
This discrepancy does not affect our main conclusions.
For the sake of the present paper, the key result is that the \modelname{Exp}{2} model successfully reproduces the age independence of stellar population metallicities.
This success is most obvious in [O/H], but the model still predicts some evolution toward high [Fe/H] at $R \lesssim 5$ kpc across the age range shown in Figure \ref{fig:allmodels-gradxh-agebins}.
This prediction arises in all of our models due to SN Ia time delays slowing the approach to equilibrium (see the dotted curves in the top row of Figure \ref{fig:allmodels-evol}).
\par
We have also construcated an \modelname{Exp}{3} model with $y / Z_\odot = 3$ (not shown) and a higher $\eta_\odot = 2.4$ to compensate for the larger yield normalization.
As expected, this parameter choice leads to a more rapid approach to equilibrium, less evolution between the age bins in the stellar metallicity gradient, and better agreement with APOGEE.
We do not explore this model in detail because the high yield normalization appears incompatible with the empirical constraints laid out by \citet{Weinberg2024}.
The differences can also be compensated by non-zero metallicity accretion and/or an overall metallicity dependence to SN rates (see discussion in Section \ref{sec:discussion:yields:norm}).
Such prescriptions may also help the \modelname{Exp}{2} reach [Fe/H] $\gtrsim$ 0.3 at $R \lesssim 5$ kpc earlier, improving its agreement with the data in Figure \ref{fig:allmodels-gradxh-agebins}.

\subsection{The Equilibrium Gradient}
\label{sec:results:eq-gradient}

\begin{figure*}[!ht]
\centering
\includegraphics[scale = 0.85]{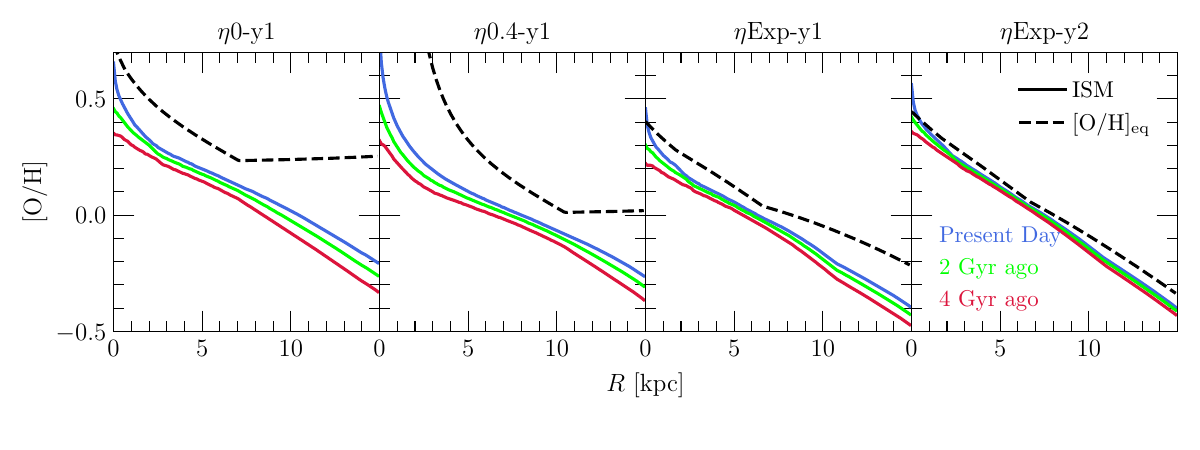}
\caption{
Evolution in gas-phase O abundance gradients in comparison to their equilibrium gradients.
Solid lines show the abundances at the present day and at two recent snapshots predicted by each of the four GCE models in our primary set (see Table \ref{tab:gcemodels} and discussion in Section \ref{sec:gce}).
Black dashed lines mark the equilibrium gradients in each model (see Equation \ref{eq:zoeq-waf17}).
{\bf Summary}: The \modelname{Exp}{2} model is very near the equilibrium metallicity gradient, while other models do not reach this state within the disk lifetime.
}
\label{fig:eq-gradient-snapshots}
\end{figure*}

\begin{figure*}
\centering
\includegraphics[scale = 0.85]{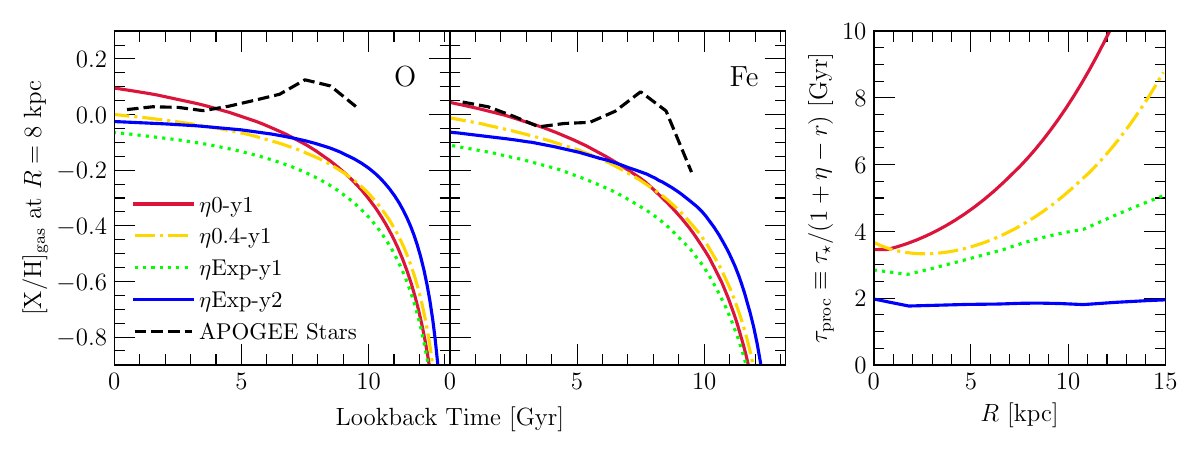}
\caption{
The key difference between our GCE models, marked according to the legend in the left panel.
{\bf Left/Middle}: The O (left) and Fe (middle) abundances at $R = 8$ kpc as a function of time as predicted by our primary set of GCE models (see Table \ref{tab:gcemodels} and discussion in Section \ref{sec:gce}).
For comparison, we plot the values inferred from our linear regressions reported in Table \ref{tab:regressions}, which are identical to the red (O) and blue (Fe) lines in the lower right panel of Figure \ref{fig:gradxh-fixedage} (see discussion in Section \ref{sec:empirical:evolution}).
{\bf Right}: The processing timescale of the ISM (see Equation \ref{eq:tauproc}) as a function of radius at the present day.
{\bf Summary}: Reaching high abundances early in the disk lifetime requires short processing timescales, meaning that a typical fluid element does not remain in the ISM for long ($\lesssim$$1 - 2$ Gyr) before being incorporated into new stars or ejected in an outflow.
}
\label{fig:tauproc-demo}
\end{figure*}

In this subsection, we discuss the physical origin of radially dependent equilibrium metallicities in the ISM by comparing between models.
In Section \ref{sec:results:sfr-per-ifr} below, we demonstrate that the decline in $Z_\text{eq}$ with radius tracks a decline in the ratio of star formation to accretion, $\dot{\Sigma}_\star / \dot{\Sigma}_\text{in}$, on long timescales.
This result can be understood as $\dot{\Sigma}_\star$ setting the metal production rate while $\dot{\Sigma}_\text{in}$ sets the rate at which fresh H is added to the ISM.
This relationship arises in the limit that infalling material is zero metallicity, but it should remain accurate as long as accretion is sufficiently below the ISM metallicity.
% \begin{subequations}\begin{align}
% \nabla_\text{eq} &\equiv \frac{
%     \partial \log_{10} (Z_\text{eq} / Z_\odot)
% }{
%     \partial R
% }
% \\
% &\rightarrow \frac{1}{\ln 10} 
% \end{align}\end{subequations}
% where $\nabla_\text{eq}$ is the equilibrium gradient.
\par
Figure \ref{fig:eq-gradient-snapshots} shows the predicted O abundance in the ISM as a function of radius in two recent snapshots and at the present day.
We also plot the equilibrium abundance $\onh{O}_\text{eq}$ as defined by Equation \refp{eq:zoeq-waf17} (using the values of $\tau_\star$ at the final snapshots).
In the \modelname{0}{1} and \modelname{0.4}{1} models, the equilibrium gradient flattens at the radii beyond which our parameter calibration assumes $\tau_\text{sfh} = 200$ Gyr (see Figure \ref{fig:calibration}, Equation \ref{eq:zoeq-waf17}, and discussion in Section \ref{sec:gce:calibration}).
Our fiducial equilibrium model, \modelname{Exp}{2}, shows little to no evolution across this time interval, because it approaches the steady state early in the disk lifetime (see discussion below).
The other models in our primary set exhibit more significant evolution in the ISM abundances over this time interval.
As expected, each model tends to predict faster evolution in the ISM metallicity when it is further below the local equilibrium abundance.
% The \modelname{0.4}{1} model is an interesting case, because $Z_\text{eq} < 0$ within $R \lesssim 6$ kpc.
% Such a prediction is not unphysical; the denominator of Equation \refp{eq:zoeq-waf17} is negative due to a sharply declining SFH (i.e. low $\tau_\text{sfh}$; see Figure \ref{fig:allmodels-evol}).
% In this case, the approach to equilibrium described by Equation \refp{eq:zo-approach-zoeq} below is inaccurate.
% Since these regions can never possibly reach an equilibrium state in this model, the ISM metallicity simply increases monotonically with time and will continue to do so until $t \rightarrow \infty$ (see discussion in Appendix \ref{sec:calibration}).
\par
The left and middle panels of Figure \ref{fig:tauproc-demo} show the predicted O and Fe abundances in the ISM at $R = 8$ kpc as a function of lookback time.
The models differ most importantly in the time dependence of the abundance evolution.
The \modelname{Exp}{2} model predicts metal abundances to change only minimally over the last $\sim$$8 - 10$ Gyr of disk evolution.
Other models in our primary set predict the ISM metallicity to increase more gradually until the present day.
This difference is more obvious in \onh{O} than \onh{Fe}, because alpha elements reach the local equilibrium abundance earlier than iron-peak elements due to the SN Ia DTD \citep{Weinberg2017b}.
\par
For comparison, we include the values of \onh{O} and \onh{Fe} determined by our linear regressions applied to mono-age populations in Figure \ref{fig:gradxh-fixedage} (reported in Table \ref{tab:regressions}).
Our models predict the ISM metallicity at $R = 8$ kpc to be marginally ($\sim$$0.1$ dex) lower than observed in stars, with the exception of the \modelname{0}{1} model $\lesssim 4$ Gyr ago.
This systematic offset could be corrected by a similar ($\sim$0.1 dex) increase in our metal yields in all models, which would be comfortably within the range allowed by the uncertainties in stellar nucleosynthesis (see discussion in Section \ref{sec:gce}).
Alternatively, the strength of outflows could be lowered by a similar amount (see dicussion in Section \ref{sec:discussion:yields:norm}).
Radial migration should also slightly increase characteristic abundances by $\sim$$0.1 - 0.2$ dex (see Figure \ref{fig:oh-dfs-with-vs-no-migration} and discussion in Section \ref{sec:discussion:migration:mdfs}) and may be related to the ``bump'' seen in the stellar abundances at age $\approx 7 - 8$ Gyr.
Although each model provides a reasonable explanation for the metallicities observed in young stars, the data indicate that the age dependence is minimal at most up to populations as old as $\sim$$8 - 10$ Gyr.
This realization was also highlighted recently by \citet{Gallart2024}.
The stability of ISM abundances predicted by the \modelname{Exp}{2} model plays a key role in holding stellar metallicities fixed between mono-age populations.
% The agreement is not perfect, but our \modelname{Exp}{2} model provides a better explanation than our other models for this weak dependence on age.
\par
In the right panel of Figure \ref{fig:tauproc-demo}, we plot the ``processing timescale\footnote{
    In \citet{Weinberg2017b} and \citet{Johnson2021}, we refer to this timescale as the ``depletion time.''
    We change nomenclature in this paper, because this term has other meanings in the literature.
},'' defined as
\begin{equation}
\tau_\text{proc} \equiv \frac{\tau_\star}{1 + \eta - r},
\label{eq:tauproc}
\end{equation}
as a function of radius at the present day.
This quantity describes the average time interval that a given ISM fluid element will remain present before being incorporated into new stars or ejected in an outflow.
In a one-zone model with a constant SFH, \citet{Weinberg2017b} demonstrate that the ISM reaches the equilibrium alpha element abundance on this timescale:
\begin{equation}
Z_\text{O}(t) = Z_\text{O,eq} \left(1 - e^{-t / \tau_\text{proc}}\right).
\label{eq:zo-approach-zoeq}
\end{equation}
The corrective factor for most other forms of the SFH is small, except in the ``gas starved'' regime where accretion falls far below the star formation and outflow rates.
% with our \modelname{0.4}{1} being a noteworthy exception (see discussion in Appendix \ref{sec:calibration}).
The processing timescale can be qualitatively understood as setting the rate of approach to equilibrium, because the ISM loses some memory of its initial chemical composition every time it is effectively replaced with new matter.
Consequently, an equilibrium arises after a handful of ``generations'' of baryons in the ISM, the duration of which is set by $\tau_\text{proc}$.
\par
The important connection illustrated by Figure \ref{fig:tauproc-demo} is that the \modelname{Exp}{2} model reaches high abundances more quickly than the other models because it predicts the shortest processing timescales.
There is a relationship between the value of $\tau_\text{proc}$ and the slope of \onh{X} as a function of time in the ISM, with shorter processing timescales corresponding to more constant abundances.
The trend in $\tau_\text{proc}$ with radius is sometimes non-monotonic, since inefficient star formation slows down ISM processing at large radii but strong outflows speed it up (see Equation \ref{eq:tauproc}).
% For the sake of qualitatively understanding equilibrium chemical evolution, the important connection illustrated by Figure \ref{fig:tauproc-demo} is that models predicting longer processing timescales have slower enrichment histories.
% The \modelname{Exp}{2} model has the shortest $\tau_\text{proc}$, so it reaches solar abundances at $R = 8$ kpc first.
% The trend in $\tau_\text{proc}$ with radius is non-monotonic in some models, because inefficient star formation slows down processing of the ISM at large radii, but strong outflows speed it up (see Equation \refp{eq:tauproc}).
% Each model has at least one break in the relation, which corresponds to a radius at which our parameter calibration procedure changes strategy (see discussion in Section \ref{sec:gce:calibration} and Appendix \ref{sec:calibration}).
% The $\eta \propto e^R$ models switch from searching for values of $\tau_\text{sfh}$ to $\tau_\text{rise}$ at $R \approx 6$ kpc, while the $\eta = $ constant models adopt a fail-safe solution at $R \gtrsim 7$ kpc and $R \gtrsim 10$ kpc.

\subsection{The Ratio of Star Formation to Accretion}
\label{sec:results:sfr-per-ifr}

\begin{figure}
\centering
\includegraphics[scale = 0.9]{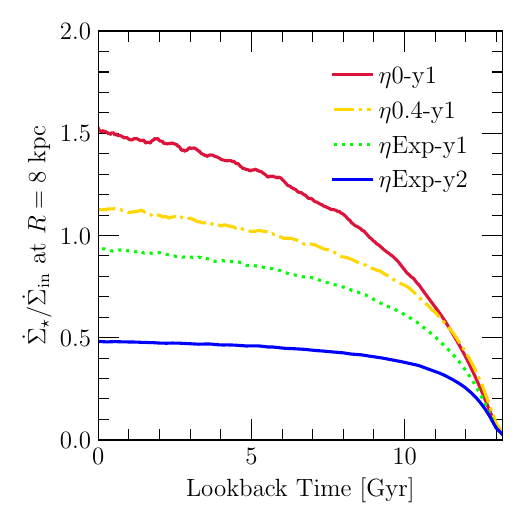}
\caption{
The star formation per unit infall at $R = 8$ kpc as a function of lookback time predicted by our primary set of GCE models.
For visual clarity, we have box-car smoothed the trend with a window width of $200$ Myr.
The ISM reaches the local equilibrium abundance only after this ratio has reached some constant value (see discussion in Section \ref{sec:results:sfr-per-ifr}).
{\bf Summary}: This ratio quickly becomes flat in the \modelname{Exp}{2} model due to a short processing timescale (see also Figure \ref{fig:tauproc-demo}).
}
\label{fig:sfr-per-ifr}
\end{figure}

\begin{figure*}
\centering
\includegraphics[scale = 0.95]{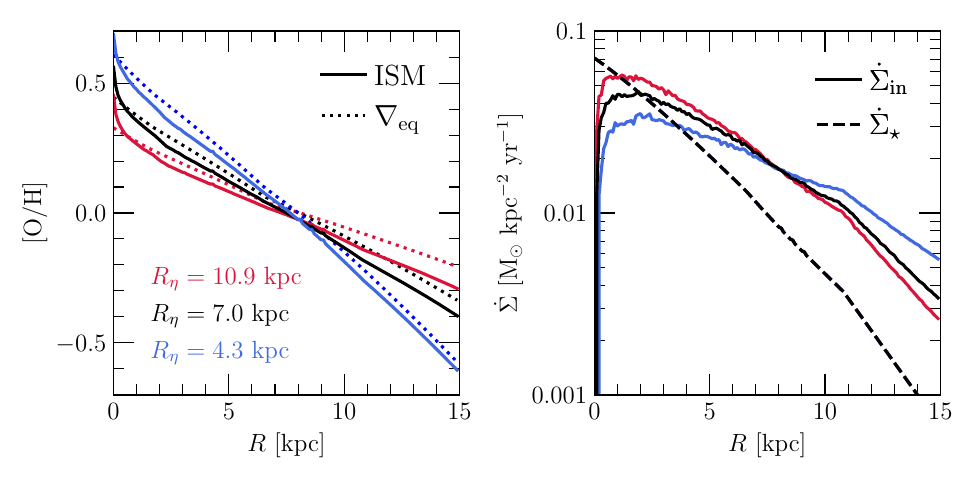}
\caption{
% {\color{red}
% I think the equilibrium gradients plotted here is just a simple one that slaps the slope on with [O/H] = 0 at $R = 8$ kpc.
% Probably better to actually compute the equilibrium abundances taking into account $\tau_\text{sfh}$ and the current $\tau_\star$ as functions of radius.
% Likely has something to do with why the blue model slightly overshoots the equilibrium abundance at large $R$ and seems interestingly below at small $R$.
% }
The relationship between the ratio of star formation per unit infall $\dot{\Sigma}_\star / \dot{\Sigma}_\text{in}$ and the slope of the equilibrium metallicity gradient $\nabla_\text{eq}$.
Models are color coded by the value of $R_\eta$, which controls the slope of the equilibrium gradient (see discussion in Section \ref{sec:gce:primary-set}), according to the legend in the left panel.
{\bf Left}: The ISM [O/H] abundance at the present day (solid) and the equilibrium gradients (dotted).
{\bf Right}: The surface densities of star formation (dashed) and infall (solid) at the present day as a function of Galactocentric radius.
The SFRs are the same in all three models.
{\bf Summary}: The equilibrium gradient traces variations in the time-averaged ratio of star formation per unit infall $\dot{\Sigma}_\star / \dot{\Sigma}_\text{in}$ with radius (see discussion in Section \ref{sec:results:sfr-per-ifr}).
}
\label{fig:sfr-ifr-gradslope}
\end{figure*}

Figure \ref{fig:sfr-per-ifr} shows the ratio of star formation per unit infall as a function of time at $R = 8$ kpc in our primary set of GCE models (see discussion in Section \ref{sec:gce:primary-set}).
% All models except \modelname{Exp}{2} predict $\dot{\Sigma}_\star / \dot{\Sigma}_\text{in} = 1.1 - 1.3$ at the present day.
The maximum allowed value for a constant SFH is $1 / (1 + \eta - r)$ (see Equation \ref{eq:sfr-per-ifr-gceparams} below), so each model asymptotically approaches a different value.
The \modelname{Exp}{2} model reaches its maximum early in the disk lifetime, while other models follow suit on longer timescales.
\par
There is information in both the time dependence of this ratio and its final value.
In combination with the scale of stellar yields, the final value sets the local equilibrium abundance (see discussion below).
The \modelname{Exp}{2} model predicts $\dot{\Sigma}_\star / \dot{\Sigma}_\text{in}$ to be lower than the other models, but it reaches similar metallicities because the scale of stellar yields is also a factor of 2 higher.
Once $\dot{\Sigma}_\star / \dot{\Sigma}_\text{in}$ becomes time independent, the ISM reaches the local equilibrium metallicity.
The \modelname{Exp}{2} model reaches the equilibrium state so early in the disk lifetime because this ratio is nearly constant for $\sim$10 Gyr.
\par
It is straightforward to demonstrate this connection mathematically.
From Equation \refp{eq:dot-sigma-gas} in Appendix \ref{sec:calibration}, it follows algebraically that $\dot{\Sigma}_\star / \dot{\Sigma}_\text{in}$ can be expressed in terms of our GCE parameters as
\begin{equation}
\frac{\dot{\Sigma}_\star}{\dot{\Sigma}_\text{in}} = \left[
1 + \eta - r + \tau_\star \frac{\dot{\Sigma}_g}{\Sigma_g}
\right]^{-1},
\label{eq:sfr-per-ifr-gceparams}
\end{equation}
where we have also substituted in the SFE timescale $\tau_\star \equiv \Sigma_g / \dot{\Sigma}_\star$.
The rate of change in the O abundance (see Equation \ref{eq:dot-z-o}) can then be expressed as
\begin{equation}
\dot{Z}_\text{O} = \frac{\ycc{O}}{\tau_\star} -
\frac{Z_\text{O}}{\tau_\star} \left(
\frac{\dot{\Sigma}_\text{in}}{\dot{\Sigma}_\star}\right).
\end{equation}
Furthermore, if $\tau_\star$ is constant and the late-time SFH declines exponentially, then $\dot{\Sigma}_g / \Sigma_g = -1 / \tau_\text{sfh}$, and the quantity in square brackets in Equation \refp{eq:sfr-per-ifr-gceparams} is equivalent to the denominator of Equation \refp{eq:zoeq-waf17} for $Z_\text{O,eq}$.
The equilibrium abundance can then be expressed as
\begin{equation}
Z_\text{O,eq} = \ycc{O}
\frac{\dot{\Sigma}_\star}{\dot{\Sigma}_\text{in}}.
\label{eq:zoeq-sfr-per-ifr}
\end{equation}
We emphasize that this expression is only accurate at $Z_\text{ISM} \approx Z_\text{eq}$.
Otherwise, $\dot{\Sigma}_\star / \dot{\Sigma}_\text{in}$ does not reflect the equilibrium state.
Equation \refp{eq:zoeq-waf17} is a more reliable expression for $Z_\text{eq}$, because it is written in terms of the input GCE parameters.
\par
The equilibrium gradient slope then follows by taking the logarithm of Equation \refp{eq:zoeq-sfr-per-ifr} and differentiating with respect to radius:
\begin{equation}
\nabla_\text{eq} = \frac{1}{\ln 10} \left[\frac{
    \partial \ln \dot{\Sigma}_\star
}{
    \partial R
} - \frac{
    \partial \ln \dot{\Sigma}_\text{in}
}{
    \partial R
}
\right].
\label{eq:eq-gradient}
\end{equation}
This expression neglects changes in the mass fraction of H, because it is typically a small correction ($\sim$$0.01$ dex).
Due to the short lifetimes of massive stars \citep[e.g.,][]{Larson1974, Maeder1989, Henry2000}, this relation is most accurate for alpha elements.
The broad nature of the SN Ia DTD introduces an additional sensitivity to the shape of the SFH in the equilibrium abundance of Fe \citep{Weinberg2017b}, which is likely related to why $\grad{Fe}$ is slightly steeper than $\grad{O}$ (see Figure \ref{fig:gradients} and Table \ref{tab:regressions}).
\par
To demonstrate this relationship in action, we compare the \modelname{Exp}{2} model to the \modelname{Exp}{2}-steep and \modelname{Exp}{2}-shallow variations (see discussion in Section \ref{sec:gce:variations}).
Each of these models has the same value of $\eta_\odot$, but with different scale radii $R_\eta$ describing how quickly the mass loading factor increases with radius.
The left panel of Figure \ref{fig:sfr-ifr-gradslope} shows the present-day O abundance in the ISM as a function of radius in comparison to the equilibrium gradients predicted by these models (computed by evaluating Equation \ref{eq:zoeq-waf17} at all radii).
Each model is near its equilibrium state at the present day by construction, but the equilibrium gradients have different slopes.
\par
The right panel of Figure \ref{fig:sfr-ifr-gradslope} shows the predicted star formation and accretion rates as a function of radius at the present day.
The SFR is the same between all three models, because it is specified as an input (see discussion in Section \ref{sec:gce}).
The inferred accretion rates differ as a consequence of changes in the mass loading factor.
Relative to the fiducial model, an increase in the mass loading factor is accompanied by a larger accretion rate, since more mass is required to fuel star formation and fulfill the stellar mass budget of the Galactic disk when more ISM material is ejected.
The \modelname{Exp}{2}-shallow model at $R < 8$ kpc and the \modelname{Exp}{2}-steep model at $R > 8$ kpc are examples of this prediction.
Conversely, \modelname{Exp}{2}-shallow at $R > 8$ kpc and \modelname{Exp}{2}-steep at $R < 8$ kpc have lower accretion rates due to weaker mass loading than the fiducial model.
\par
There is a clear inverse relation in the local accretion rate and metallicity at fixed SFR, which is a direct consequence of this connection.
These \modelname{Exp}{2}-shallow and \modelname{Exp}{2}-steep models predict higher (lower) \onh{O} than the fiducial model if the accretion rate is lower (higher).
The effect of mass loaded outflows on metal enrichment rates, as described by \citet{Weinberg2017b} and supported by Figure \ref{fig:sfr-ifr-gradslope}, is to remove metal-rich material from the ISM and replace it with metal-poor gas through accretion.
This ongoing dilution lowers the equilibrium abundance.
If this process happens in a radially dependent manner, then a radial gradient in the equilibrium metallicity arises.
We note that we have found similar predictions when holding the accretion rate $\dot{\Sigma}_\text{in}$ fixed between models, in which case the surface density of star formation $\dot{\Sigma}_\star$ varies as a consequence of different choices in $\eta$.

\subsection{Perturbations due to Major Mergers}
\label{sec:results:perturbative}

% \begin{figure}
% \centering
% \includegraphics[scale = 0.95]{eta-sudden-change.pdf}
% \caption{
% {\color{red}
% It is probably worth cutting the \modelname{evolExp}{3} model entirely as it does not seem to be discussed to significant depth with an updated version of the text.
% It can be stated in words somewhere that changing $R_\eta$ results in a tilt of the gradient as expected.
% }
% The impact of building a sudden change in $R_\eta$ at $t = 8$ Gyr into our \modelname{Exp}{2} model (see Table \ref{tab:gcemodels} and discussion in Section \ref{sec:gce}).
% {\bf Top}: The mass loading factor $\eta$ as a function of radius before (solid) and after (dashed) the perturbation.
% {\bf Bottom}: The [O/H] abundance in the ISM at the present day (black) and at a handful of snapshots immediately following the parameter change (lines color coded according to the legend).
% {\bf Summary}: Parameter changes lead to changes in the equilibrium abundance with time, which can lead to shifts in the slope of the metallicity gradient.
% Though we do not find any obvious signs of such events in our sample, the literature has seen some recent arguments in favor of evolution in its slope (see discussion in Section \ref{sec:discussion:literature}).
% }
% \label{fig:eta-sudden-change}
% \end{figure}

\begin{figure}
\centering
\includegraphics[scale = 0.92]{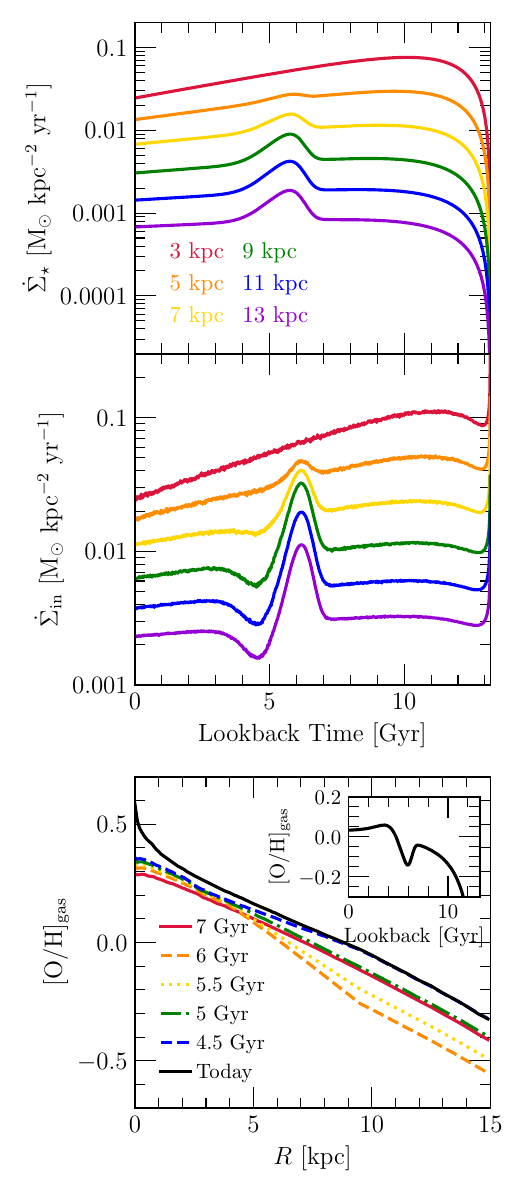}
\caption{
Our \modelname{Exp}{2}+burst GCE model (see Table \ref{tab:gcemodels} and discussion in Section \ref{sec:gce:variations}).
{\bf Top/Middle}: The input SFH (top) and predicted infall history (middle) at a selection of six Galactocentric radii, color coded according to the legend in the top panel.
{\bf Bottom}: [O/H] in the ISM as a function of Galactocentric radius, marked by the lookback time of the snapshot according to the legend.
{\bf Inset}: [O/H] in the ISM at $R = 8$ kpc as a function of lookback time.
{\bf Summary}: This model predicts the metallicity gradient to vary with a scale and timing that is broadly consistent with the $\sim$$4 - 6$ Gyr populations in our sample (see Figure \ref{fig:gradxh-fixedage} and discussion in Section \ref{sec:empirical:evolution}).
}
\label{fig:outerburst}
\end{figure}

% As additional explorations of the parameter space, we discuss the \modelname{evolExp}{3} and \modelname{Exp}{2}+burst models in this section (see discussion in Section \ref{sec:gce:variations}).
% The \modelname{evolExp}{3} model is defined by a sudden change in $R_\eta$ from a value of $7$ kpc to $21.7$ kpc.
% As a result, the mass loading factor becomes suddenly larger at small radii and lower at large radii.
% The equilibrium gradient pivots around the solar abundance at $R = 8$ kpc to a new slope of $\nabla \onh{O}_\text{eq} = -0.02$ kpc$^{-2}$.
% \par
% Figure \ref{fig:eta-sudden-change} shows the initial and final scalings of $\eta$ with radius alongside snapshots of the O abundance in the ISM immediately following the change in $R_\eta$ at $t = 8$ Gyr.
% As expected, the ISM metallicity shifts gradually from its initial equilibrium state to the new equilibrium state on $\sim$Gyr timescales.
% The observational signature of such a model is a change in the gradient slope $\nabla\onh{O}$ and $\nabla\onh{Fe}$ in mono-age stellar populations.
% We discuss this prediction further in the context of recent arguemnts in the literature in Section \ref{sec:discussion:birth-radii} below.
% \par
As an additional exploration of the parameter space, the \modelname{Exp}{2}+burst model is defined by a burst in star formation $\tau \approx 6$ Gyr ago in the central and outer disk but is otherwise the same as our \modelname{Exp}{2} model (see Equation \ref{eq:burst-sfh} and discussion in Section \ref{sec:gce:variations}).
The top and middle panels of Figure \ref{fig:outerburst} show the star formation and accretion histories at a selection of Galactocentric radii.
In the outer disk, the accretion rates increase by a factor of a few for $\sim$$1 - 2$ Gyr in order to fuel the enhanced star formation activity.
The ``jitter'' in these predictions arises due to minor, stochastic fluctuations in the rate of return of stellar envelopes back to the ISM introduced by stellar migration, which affects the inferred level of accretion required to fuel the specified level of star formation (see discussion in Section \ref{sec:results:sfr-per-ifr} above).
\par
The bottom panel of Figure \ref{fig:outerburst} shows the O abundance in the ISM as a function of radius at a handful of snapshots covering the duration of the starburst.
The ISM metallicity first decreases due to the dilution associated with the enhanced accretion rates.
Following this time period, re-enrichment from star formation elevates the gas-phase abundances back to their pre-burst values within $\sim$$2$ Gyr.
The inset shows this prediction as a function of time at $R = 8$ kpc.
The effects of this accretion induced burst of star formation can be understood as a perturbation that evolves back to the initial equilibrium state on $\sim$Gyr timescales, in line with previous theoretical arguments about dilution and re-enrichment \citep[e.g.,][]{Dalcanton2007, Johnson2020}.
\par
The parameters of the \modelname{Exp}{2}+burst model are {\it ad hoc}, chosen simply because they qualitatively explain the metallicity variations in intermediate aged populations in the outer disk seen in Figure \ref{fig:gradxh-fixedage}.
Based on the lookback time to the enhanced accretion, this event could plausibly be associated with the first pericentric passage of the Sagittarius dwarf spheroidal \citep[e.g.,][]{Law2010, RuizLara2020}.
% (Sgr dSph; e.g., \citealt{Law2010, RuizLara2020}.
It is also possible that the MW experienced a burst of star formation that was coincident but unassociated with Sagittarius.
Simulations predict that merger events fuel bursts of star formation that are centrally concentrated as opposed to localized to the outer disk due to gas flows induced by global torques \citep[e.g.,][]{Hopkins2013}.
Whether or not Sagittarius induced these large scale flows is a question of whether or not it was massive enough to do so.
However, in observations, the connection between merger events, enhanced star formation, and the radial dependence thereof is unclear on a galaxy-by-galaxy basis (see discussion in, e.g., \citealt{Pearson2019} and \citealt{Thorp2024}).
% In simulations, merger events induce gas flows due to global torques, which fuels a burst of star formation that is centrally concentrated as opposed to localized to the outer disk \citep[e.g.,][]{Hopkins2013}.
% However, it is possible that Sagittarius is not massive enough to induce these flows.
% Observationally, the connection between merger events, enhanced star formation, and the radial profile thereof is unclear on a galaxy-by-galaxy basis (see discussion in, e.g., \citealt{Pearson2019} and \citealt{Thorp2024}).
% The parameters of the \modelname{Exp}{2}+burst model are {\it ad hoc} (see discussion in Section \ref{sec:gce:variations}), so the MW may have also experienced a burst of star formation in the outer disk that was coincident but unassociated with Sagittarius.
\par
We have also explored variations in which the value of $R_\eta$ changes suddenly at some point in the disk lifetime, leading to a shift in the equilibrium gradient.
As expected, the metal abundances evolve from the original values at the time of the shift and settle onto the new equilibrium gradient on $\sim$Gyr timescales.
These models underscore the notion that the equilibrium scenario allows for changes in the slope over time, which the \modelname{Exp}{2}+burst model indicates should be expected from major mergers anyway.
In general, some level of evolution should be expected if the equilibrium is disrupted or perturbed.

\section{Discussion}
\label{sec:discussion}

\subsection{Mass-Loaded Outflows}
\label{sec:discussion:outflows}

Our most successful model is \modelname{Exp}{2} because it reaches an equilibrium metallicity gradient early in the disk lifetime.
This prediction plays a key role in reproducing the observed lack of evolution in metallicity with the ages of stellar populations across the Galactic disk (see Figure \ref{fig:allmodels-gradxh-agebins} and discussion in section \ref{sec:results}).
An outflow that ejects ISM material more efficiently with increasing Galactocentric radius is what ultimately enables the equilibrium behavior of this model (see discussion in section \ref{sec:gce:primary-set}).
% Ejection lowers the amount of time that baryons remain in the ISM, which allows the equilibrium state to be reached more quickly.
% Outflows also lower the amount of star formation per unit accretion, $\dot{\Sigma}_\star / \dot{\Sigma}_\text{in}$, which in turn lower the local equilibrium abundance (see discussion in section \ref{sec:results:sfr-per-ifr}).
\par
Many previous GCE models of the MW disk omit outflows, using $\eta = 0$ and implicitly assuming that the gravitational potential is too strong to launch a substantial wind \citep[e.g.,][]{Minchev2013, Minchev2014, Spitoni2019, Spitoni2021, Palla2020, Palla2022, Palla2024, Gjergo2023}.
Some models deposit a portion of stellar yields directly into an outflow, but ambient ISM material is not ejected \citep[e.g.,][]{Schoenrich2009a, Chen2023}.
These arguments are often based on hydrodynamic simulations in which ejected metals are re-accreted on short timescales ($\lesssim$$100$ Myr; \citealt{Spitoni2008, Spitoni2009}) near the SN that produced them ($\lesssim$$0.5$ kpc; \citealt{Melioli2008, Melioli2009, Hopkins2023}).
However, feedback is both extremely uncertain and sensitive to the implementation of the simulation \citep[e.g.,][]{Li2020, Hu2022}.
Mass-loaded outflows emerge in other simulations of MW-like galaxies with different prescriptions \citep[e.g.,][]{Brook2011, Gutcke2017, Nelson2019, Peschken2021, Kopenhafer2023}.
\par
From a purely empirical perspective, outflows with substantial mass loading have not yet been observed from MW-like galaxies, but the predicted column densities are below the detection thresholds of current instruments (see discussion in, e.g., the reviews by \citealt{Veilleux2020} and \citealt{Thompson2024}).
To date, evidence of mass-loading is largely limited to nearby starbursting systems (e.g., M82, \citealt{Lopez2020}; NGC 253, \citealt{Lopez2023}; Mrk 1486, \citealt{Cameron2021}).
However, results in recent years suggest that galaxies forming stars under more MW-like conditions may still eject ISM material in outflows.
\citet{ReichardtChu2022a} report the detection of outflows at 500 pc resolution in regions of low star formation surface density in otherwise strongly star forming galaxies.
\citet{Avery2022} also argue based on NaI lines that at least some disk galaxies have much larger neutral outflows than ionized outflows.
Feedback-driven outflows have also been observed across cosmic time (locally: \citealt{Chen2010, ReichardtChu2022a, ReichardtChu2022b, Avery2021, Avery2022}; at $z \sim 1$: \citealt{Rubin2010}; at $z \sim 2$: \citealt{Davies2019, Concas2022}).
\citet{Rudie2019} showed that $\sim$70\% of $z \sim 2$ galaxies with detected metal absorption contain some amount of metal-enriched gas whose line-of-sight velocity exceeds the 3D escape velocity.
Based on these results, it is not unlikely that the MW has ejected a substantial outflow at some point in its evolutionary history, even if it is only producing fountain flows at the present day.
\par
Our prescription for outflows in this paper is {\it ad hoc}, however, since we chose the $\eta \propto e^R$ formalism because it would produce an equilibrium gradient of approximately the correct shape (see discussion in section \ref{sec:gce:primary-set}).
Observationally, mass loading appears to increase with the surface density of both stars and star formation \citep[e.g.,][]{ReichardtChu2024}, which would suggest that $\eta$ should be highest in the inner disk as opposed to the outer disk.
If our $\eta \propto e^R$ prescription is inaccurate, then a potential alternate origin for equilibrium chemical evolution in the MW is radial gas flows \citep[e.g.,][]{Lacey1985, Portinari2000, Spitoni2011, diTeodoro2021, Trapp2022, DuttaChowdhury2024}.
These flows are thought to be directed inward because the angular momentum of accreting material is generally lower than that of the disk \citep{Bilitewski2012}.
This process may play a role in regulating the ISM abundance through dilution by incorporating low metallicity gas from larger radii into the ambient ISM in a given region of the disk.
We plan to investigate the role of radial gas flows in future work, along with more empirically motivated prescriptions for $\eta$.
For any model of the origin of the metallicity gradient in the MW, the lack of evolution in stellar metallicities across the last $\sim$9 Gyr is an important empirical constraint.

% \subsubsection{The Normalization of Stellar Yields}
% \label{sec:discussion:gas-flows:yield-norm}
\subsection{Stellar Yields}
\label{sec:discussion:yields}

\subsubsection{The Normalization}
\label{sec:discussion:yields:norm}

\begin{figure}
\centering
\includegraphics[scale = 0.9]{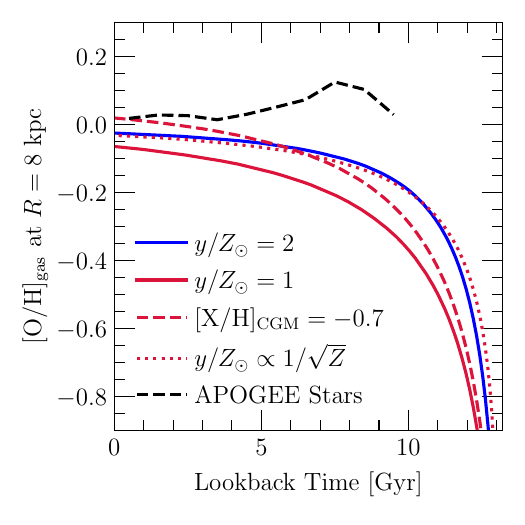}
\caption{
Variations in the \modelname{Exp}{1} model (all shown in red), which are intended to quicken its approach to equilibrium (see discussion in section \ref{sec:discussion:yields:norm}), with the \modelname{Exp}{2} model shown in blue.
All curves show the ISM \onh{O} abundance as a function of lookback time at $R = 8$ kpc.
The red dashed line shows a model in which the metallicity of accreting material is initially metal free but approaches $\onh{O}_\text{CGM} \approx -0.7$.
The red dotted line shows a case in which we let all SN yields increase toward low metallicity with a $1 / \sqrt{Z}$ dependence based on recent arguments about SN event rates.
The black dashed line shows the [O/H] values inferred from our linear regressions reported in Table \ref{tab:regressions} and is identical to the red line in the lower right panel of Figure \ref{fig:gradxh-fixedage}.
% The black dashed line is identical to the left panel of Figure \ref{fig:tauproc-demo}.
{\bf Summary}: Non-zero metallicity accretion, metallicity dependent SN rates, or some combination thereof can account for the differences of a factor of $\sim$2 in metal yields within the age range we focus on in this paper.
}
\label{fig:oh-ism-variants}
\end{figure}

For many purposes in GCE, there is a degeneracy between stellar yields and the strength of outflows \citep[e.g.,][]{Cooke2022, Johnson2023b, Sandford2024}.
We demonstrated in section \ref{sec:results} that strong outflows lower the processing timescale of the ISM, enabling an equilibrium gradient to be reached early in the disk lifetime.
This difference is the primary reason why \modelname{Exp}{1} and \modelname{Exp}{2} do not predict similar enrichment histories, despite both reaching the abundances observed in the ISM at the present day.
Simply ejecting metals in a hot outflow without much mass loading leads to a longer processing timescale and thus a slow approach to equilibrium.
However, a potential problem with the \modelname{Exp}{2} model is that the yields are large compared to the empirical estimates from \citet{Weinberg2024}, which are anchored to the mean Fe yield of CCSNe measured by \citet{Rodriguez2023}.
We therefore investigate two variations of the \modelname{Exp}{1} model in this section to see if a faster approach to equilibrium can be achieved.
% Both variations should be understood as deliberately simplified investigations as opposed to complete tests of their respective hypotheses.
We focus on the predicted O abundances because the results for Fe are similar.
\par
In our first variation, the metallicity of the circumgalactic medium (CGM) is initially zero but follows a $1 - e^{-t / \tau}$ dependence, reaching a value of [O/H]$_\text{CGM} =$ [Fe/H]$_\text{CGM} = -0.7$ on a $\tau = 3$ Gyr timescale.
Neither the metallicity of accreting material nor its evolution with cosmic time are well constrained.
Between redshift $z \sim 0$ and $\sim$$1$, the CGM metallicity is observed to vary by over an order of magnitude between sightlines in individual halos \citep[e.g.,][]{Zahedy2019, Zahedy2021, Cooper2021, Haislmaier2021, Kumar2024, Sameer2024}.
These results indicate that the CGM is not well-mixed, so pinning down a mass weighted metallicity is challenging.
By extension, how the CGM composition varies across cosmic time is similarly unconstrained.
Future constraints may show that this model's prescription with only two free parameters is inaccurate.
\par
% {\color{red}
% Accretion instead remains at zero metallicity in our second variation, but the ISM has an initial abundance of [O/H] = [Fe/H] = $-0.7$.
% Since our models are best constrained by thin disk stellar populations, it is not necessarily the case that they should start at $Z = 0$.
% In principle, there should have been some metals left over from the halo and thick disk formation epochs.
% }
In our second variation, we let the yields of all elements follow a $1 / \sqrt{Z}$ dependence on metallicity.
\citet{Johnson2023c} argue that this dependence, which follows the close binary fraction observed in APOGEE \citep{Badenes2018, Moe2019}, can explain the high SN Ia rates observed in dwarf galaxies \citep[e.g.,][]{Brown2019, Wiseman2021}.
\citet{Pessi2023} showed that CCSN rates also follow a strong inverse dependence on metallicity.
Their inferred scaling is stronger than $1 / \sqrt{Z}$, so it would possibly be more accurate to use different scalings for different SN types.
However, the key question is whether or not higher overall event rates at low metallicity can make up the difference between our \modelname{Exp}{1} model and the abundances observed in old stellar populations.
\par
Figure \ref{fig:oh-ism-variants} compares the predicted [O/H] abundances at $R = 8$ kpc as a function of time between these variations, our \modelname{Exp}{1} and \modelname{Exp}{2} models, and the mode of the MDF in mono-age
The differences between the model predictions and the stellar abundances in APOGEE could be due to radial migration, but could also be compensated by slight modifications to our stellar yields or the strength of outflows (see discussion in section \ref{sec:results:eq-gradient}).
% Both variations significantly increase the metallicities of old stellar populations relative to \modelname{Exp}{1} and more closely resembles the predictions of \modelname{Exp}{2}.
Our $Z$-dependent SN rate model matches the \modelname{Exp}{2} model almost exactly.
The metal-rich accretion model evolves more slowly, but it still reaches [O/H] $\approx -0.2$ by a lookback time of $\sim$9 Gyr.
Despite their simplified nature, both variations are in broad agreement with the \modelname{Exp}{2} model across the range of stellar ages that we have focused on in this paper.
Metal-rich accretion and metallicity dependent SN rates are not mutually exclusive, so both effects could play a role in raising the metallicities of old stellar populations.
It is also possible that the thin disk simply started with a non-zero metallicity, since it was preceded by the halo and thick disk epochs.
Our models begin with no mass in the ISM by construction, so we refrain from investigating these models as they would require re-parameterizing the evolutionary history of our fiducial model.
Nonetheless, building in an initial metal abundance should also act to raise the metallicities of old stellar populations relative to a model that starts with no metals.

\subsubsection{Empirical Constraints}
\label{sec:discussion:yields:constraints}

If the equilibrium scenario is accurate, then trends in abundance ratios should trace trends in stellar yields.
This implication can be seen by inspecting Equation \refp{eq:zoeq-waf17} and the equivalent for Fe from section 2.4 of \citet{Weinberg2017b}.
The abundance ratio of some element X relative to O or Mg can be expressed as:
\begin{subequations}\begin{align}
\text{[X/$\alpha$]}_\text{eq} &=
\log_{10} \left(
\frac{Z_\text{X,eq}}{Z_{\alpha,\text{eq}}}
\right) - \log_{10} \left(
\frac{Z_{\text{X},\odot}}{Z_{\alpha,\odot}}
\right)
\\
&= \log_{10} \left(
\frac{y_\text{X}^\text{CC} + y_\text{X}^\text{Ia} \beta_\text{Ia}}{y_\alpha^\text{CC}}
\right) - \log_{10} \left(
\frac{Z_{\text{X},\odot}}{Z_{\alpha,\odot}}
\right),
\label{eq:xalpha-yield-ratio}
\end{align}\end{subequations}
where we use $\alpha$ to generically refer to O, Mg, or any element whose production is dominated by CCSNe.
Most importantly, the denominator of Equation \refp{eq:zoeq-waf17}, which depends on GCE parameters, has cancelled.
The factor $\beta_\text{Ia}$ arises due to the time delays of SNe Ia and is defined as an integral over the SFH weighted by the DTD $R_\text{Ia}$ in units of the present-day SFR:
\begin{equation}
\beta_\text{Ia} \equiv
\ddfrac{
    \int_0^{\tau_\text{disk}} 
    \dot{\Sigma}_\star(t)
    R_\text{Ia}(\tau_\text{disk} - t)
    dt
}{
    \dot{\Sigma}_\star(\tau_\text{disk})
    \int_0^\infty
    R_\text{Ia}(t)
    dt
}.
\label{eq:alpha-ia}
\end{equation}
SNe Ia require some minimum time delay (i.e., $R_\text{Ia}(t) = 0$ for small $t$), so \citet{Weinberg2017b} equivalently define this expression to omit recent star formation.
In detail, $\beta_\text{Ia}$ is only approximately constant with time, so the [Fe/H] gradient is not a perfectly stationary equilibrium.
\par
\citet{Weinberg2019} noted that trends in observed [X/Mg] ratios with [Mg/H] do not vary with spatial location in the MW.
They interpreted this result as an indication that these trends are nucleosynthetic in origin.
Variations with radius would be expected due to the inside-out growth of the disk if [X/Mg] were set by the SFH.
Equations \ref{eq:xalpha-yield-ratio} and \ref{eq:alpha-ia} indicate that a lack of variation in abundance sequences between Galactic regions is expected under the equilibrium scenario, since $\beta_\text{Ia}$ should theoretically be the same from element to element (unless different elements trace different epochs of the DTD).
\par
In principle, $\beta_\text{Ia}$ should vary with Galactocentric radius as a consequence of variations in the shape of the SFH.
This spatial dependence is likely related to why $\grad{Fe}$ is slightly steeper than $\grad{O}$ (see Figure \ref{fig:gradients} and Table \ref{tab:regressions}).
Variations in gradient slopes between elements $\grad{X}$ should therefore be sensitive to different yield ratios between CCSNe and SNe Ia, $y_\text{X}^\text{CC} / y_\text{X}^\text{Ia}$, as long as certain elements are not preferentially ejected in outflows.
We plan to pursue these potential empirical constraints on stellar yields in future work.

\subsection{Radial Migration}
\label{sec:discussion:migration}

% Radial migration collectively refers to a series of dynamical processes that impact the guiding center radii of stars, often carrying them several kpc from their birth radius \citep[e.g.,][]{Sellwood2002}.
% These processes are ubiquitous in numerical simulations of MW-like galaxies \citep[e.g.,][]{Roskar2008a, Roskar2008b, Loebman2011, Minchev2011, Bird2012, Bird2013, Kubryk2013}.
% Migration introduces significant uncertainties into models of the enrichment history of the MW, since it is generally challenging to determine the region of the disk a star originated from (see discussion in section \ref{sec:discussion:migration:birth-radii} below).

\subsubsection{The Effect on Metallicity Distribution Functions}
\label{sec:discussion:migration:mdfs}

\begin{figure*}
\centering
\includegraphics[scale = 0.87, trim = 0 -0.1in 0 0]{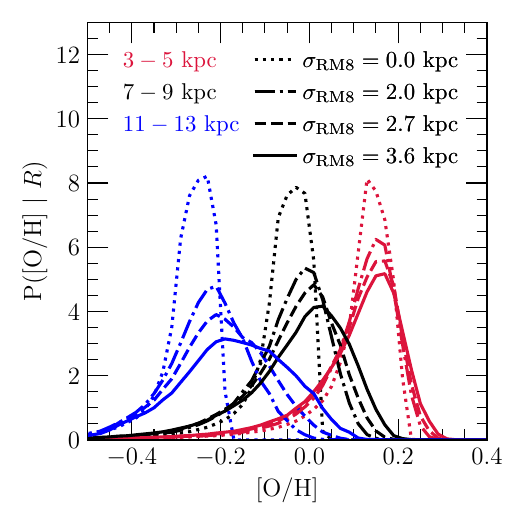}
\includegraphics[scale = 0.87]{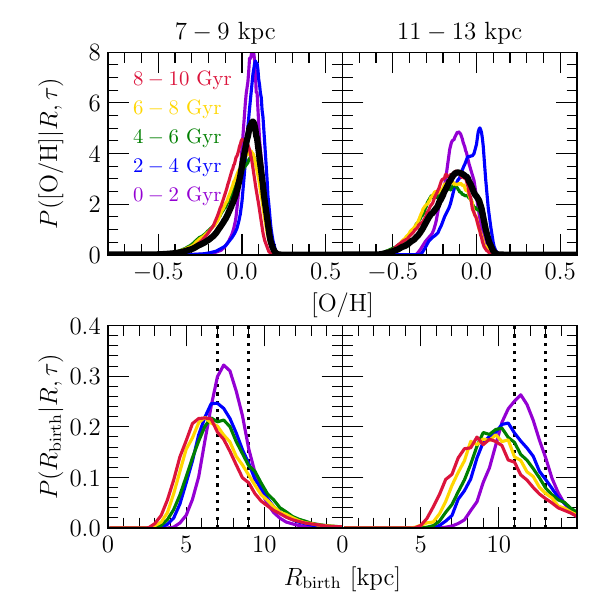}
\caption{
% {\color{red} Try a version that uses 1-kpc radial bins instead of 2-kpc.
% Correct the linestyles that are plotted (middle two are mis-matched according to the legend).
% }
The impact of radial migration on abundance distributions.
{\bf Left}: Distributions in \onh{O} in our \modelname{Exp}{2} model with different radial migration speeds $\sigma_\text{RM8}$ (see discussion in section \ref{sec:discussion:migration:mdfs}).
Lines are color coded by their 2-kpc bin in present day radius according to the legend in the top left and styled according to the legend in the top right.
{\bf Right}: Distributions in \onh{O} (top) and birth radius $R_\text{birth}$ (bottom) at $R = 7 - 9$ kpc (left) and $R = 11 - 13$ kpc (right).
Lines are color coded by stellar age according to the legend in the top left panel.
The solid black line in the top panels shows the total \onh{O} distribution.
{\bf Summary}: In the equilibrium scenario, stellar migration enhances the tails of the metallicity distribution but has only a minimal impact on the position of the peak, since the metallicity gradient is shallow compared to typical radial migration displacements (see Equation \ref{eq:oh-rbirth-dists}).
}
\label{fig:oh-dfs-with-vs-no-migration}
\end{figure*}

The left panel of Figure \ref{fig:oh-dfs-with-vs-no-migration} shows distributions in [O/H] predicted by our \modelname{Exp}{2} model in three different ranges of Galactocentric radius.
For comparison, we include variations of this model with different rates of migration.
We make this change by adjusting the parameter $\sigma_\text{RM8}$, which quantifies the width of the distribution in orbital displacement $\Delta R$ for 8 Gyr old stellar populations (see discussion in section \ref{sec:gce} and in Appendix C of \citealt{Dubay2024} for details).
Our fiducial model uses the value of $\sigma_\text{RM8} = 2.7$ kpc from \citet{Dubay2024}, which we adjust to $\sigma_\text{RM8} = 2.0$ kpc ($\sim$25\% slower) and $\sigma_\text{RM8} = 3.6$ kpc ($\sim$25\% faster; consistent with the value inferred by \citealt{Frankel2018}).
We also include a model in which we hold every star at its birth radius, which removes the effects of radial migration (i.e., $\sigma_\text{RM8} = 0$).
\par
In each model, the position of the peak of the MDF is largely unaffected at all radii.
Migration significantly enhances the wings of the distribution by contributing metal-poor stars from large radii and metal-rich stars from small radii.
This prediction is in line with the chemodynamic simulations by \citet{Grand2014}.
The stability of the MDF peak is our motivation for choosing to quantify the observed radial metallicity gradient in terms of the mode rather than the mean or median.
In our GCE models, the mode of the abundance distribution of a mono-age population at fixed radius is typically no more than $\sim$$0.1 - 0.2$ dex above the ISM abundance at the corresponding time and place.
This prediction arises in {\it all} of our models, including \modelname{0}{1} and \modelname{0.4}{1}.
The mode of the MDF becomes a slightly worse proxy for faster migration speeds, but it traces the underlying ISM abundance more closely than the mean or median in all cases because these summary statistics are sensitive to the tails of the distribution.
\par
It may seem counterintuitive that migration does not substantially affect the mode of the MDF.
At fixed present day radius, there are many more stars that have migrated outward than inward due to the radial surface density gradient.
It may be natural to expect old populations to shift toward higher metallicity as a consequence.
We do indeed see these effects in the left panel of Figure \ref{fig:oh-dfs-with-vs-no-migration} (most clearly in the $R = 11 - 13$ kpc bin with $\sigma_\text{RM8} = 3.6$ kpc), but the effect is minimal.
\par
We address this point in the right hand panel of Figure \ref{fig:oh-dfs-with-vs-no-migration}, which shows distribution in [O/H] and birth radius broken down by age in two of the radial bins from the left panel.
The shift in $P(R_\text{birth} | R, \tau)$ toward smaller radii with increasing age is comparable to the predictions of the {\tt h277} simulation by construction (for comparison, see Figure 1 in \citetalias{Johnson2021}).
The [O/H] distributions are much less sensitive to stellar age.
The two distributions are approximately related by the local slope of the ISM metallicity gradient at the corresponding lookback time according to
\begin{equation}
P(R_\text{birth} | R, \tau) \approx
P(\text{[O/H]} | R, \tau) \frac{
    \partial \text{[O/H]}
}{
    \partial R_\text{birth}
},
\label{eq:oh-rbirth-dists}
\end{equation}
but only approximately because all of these variables are correlated.
This expression indicates that the mode of the MDF is minimally affected by migration because the metallicity gradient is shallow compared to typical migration distances.
In other words, a significant excursion in radius does not correspond to a significant excursion in metallicity.
Based on our measurement of $\grad{O} = -0.06$ kpc$^{-1}$, a star that migrates $\Delta R = 3$ kpc outward is on average only $0.18$ dex more metal-rich than the stars that formed there.

% \newpage
\subsubsection{Birth Radius Inferences}
\label{sec:discussion:migration:birth-radii}

\begin{figure}
\centering
\includegraphics[scale = 0.9]{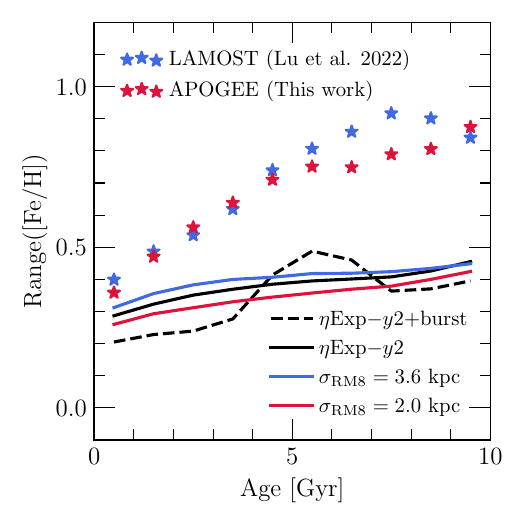}
\caption{
% {\color{red}
% Since this figure moved to the main text as opposed to the appendix where it'd be the full page width anyway, try to work the appendix into the panel and let it take up only one column.
% }
MDF width as a function of age for stars between $R = 7$ and 10 kpc as observed and predicted by our models.
The y-axis shows the difference between the 5th and 95th percentiles of the MDF.
Stars mark measurements in our sample (red) and from LAMOST by \citet[][blue]{Lu2022b}.
Black lines show the predictions by our fiducial model (solid; see discussion in section \ref{sec:gce:primary-set}) and a variation thereof with a burst of star formation in the outer disk $\sim$6 Gyr ago (dashed; see discussion in sections \ref{sec:gce:variations} and \ref{sec:results:perturbative}).
Colored lines show variations of the fiducial model with different radial migration speeds (see discussion in section \ref{sec:discussion:migration:mdfs}).
{\bf Summary}: This relationship should be substantially influenced by a burst of star formation, as argued by \citet{Lu2022b}.
However, the relationship should have a slightly different shape than $\grad{Fe}$ as a function of lookback time, and incorporating these effects may improve the \citet{Lu2022b} methodology.
Our models predict systematically narrower MDFs than observed, which could arise due to a handful of effects (see discussion at the end of section \ref{sec:discussion:migration:birth-radii}).
% {\bf Summary}: The relationship between Range(\onh{Fe}) and age should have a slightly different shape than $\grad{Fe}$ as a function of lookback time.
% Incorporating this effect would improve the \citet{Lu2022b} methodology.
}
\label{fig:lu-comparison}
\end{figure}

Recent work has sought to quantify the strength of radial migration in the Galactic disk by combining stellar ages and abundances with GCE models \citep[e.g.,][]{Frankel2018, Minchev2018}.
However, the adopted GCE models generally fall under the evolution scenario with ISM abundances that rise continually with time until the present day.
The success of of the equilibrium scenario in reproducing observations in section \ref{sec:results} therefore challenges the validity of the models these birth radius inferences are based on.
Interestingly, however, the equilibrium scenario implies that estimates of migration distances should become {\it less} model dependent.
Under the equilibrium scenario, $Z(R, \tau) \approx Z(R)$, relatively insensitive to the star's age, making metallicity a proxy for birth radius on its own.
\par
Recently, \citet{Lu2022b} constructed a new and novel method for determining birth radii.
Their method uses the width of the MDF in a mono-age population as a proxy for the metallicity gradient slope at a given lookback time.
If the decline in ISM abundances toward large radii is monotonic, then wide (narrow) MDFs should arise from steep (shallow) gradients.
They calibrated a relation between Range([Fe/H]) and $\grad{Fe}$ based on hydrodynamic simulations, after which they infer birth radii for a sample of subgiants from LAMOST \citep{Luo2015}.
\citet{Ratcliffe2023} subsequently used this method to infer birth radii for stars in APOGEE.
\par
Both \citet{Lu2022b} and \citet{Ratcliffe2023} argue that the metallicity gradient was steeper in the past based on the MDF broadening with age.
It may seem at first glance like this argument is in tension with our equilibrium scenario.
However, their argument is centered on the slope of the relation between stellar metallicities and Galactocentric radius, while ours is centered on its normalization.
In principle, the equilibrium scenario allows for changes in the slope at fixed normalization.
Such variations can arise as a consequence of major merger events or evolving Galactic properties that lead to a change in the equilibrium slope (see Figure \ref{fig:outerburst} and discussion in section \ref{sec:results:perturbative}).
\par
To compare our arguments more closely, we apply the \citet{Lu2022b} methodology to the stellar populations predicted by our GCE models (see discussion in their section 3 for full details).
To mimic the radial range of their LAMOST sample, we select only stellar populations with present day radii between 7 and 10 kpc.
We inject artificial measurement uncertainties by perturbing the ages and metallicities by random numbers drawn from normal distributions of width 0.5 Gyr and 0.03 dex, respectively.
We take Range([Fe/H]) to the difference between the 5th and 95th percentiles of the [Fe/H] distribution for a mono-age population.
We apply the same procedure to our sample from APOGEE but omit artificial uncertainties.
Although \citet{Lu2022b} do not explicitly limit their sample to $R = 7 - 10$ kpc, we make this cut to minimize differences in selection effects.
\par
Figure \ref{fig:lu-comparison} shows the results of this procedure as a function of stellar age in comparison to \citepossessivepar{Lu2022b} measurements from LAMOST.
The two surveys follow similar trends, especially considering that we are not fully taking into account the two selection functions.
Our measurements in APOGEE are also a reasonable match to \citet{Ratcliffe2023}, with some small differences in detail likely due in part to different age catalogs (they focus on the 1-$\sigma$ as opposed to 2-$\sigma$ MDF width and use a different age catalog, {\sc XGBoost}; \citealt{Miglio2021}).
We show the predicted relations from our \modelname{Exp}{2} and \modelname{Exp}{2}+burst models along with the variations that use 25\% faster and slower radial migration from section \ref{sec:discussion:migration:mdfs} above.
Despite a nearly constant slope $\grad{Fe}$, the \modelname{Exp}{2} model and its variations each predict Range([Fe/H]) to increase by $\sim$0.2 dex between young and $\sim$10 Gyr old populations.
This prediction is a consequence of radial migration.
The $R = 7 - 10$ kpc range, which this comparison focuses on, samples a broader and broader range of birth radii with increasing age.
Due to the presence of a radial abundance gradient, this region therefore also samples a broader and broader range of abundance.
\par
Our \modelname{Exp}{2}+burst model supports the argument by \citet{Lu2022b} that Range([Fe/H]) should increase and $\grad{Fe}$ should steepen briefly during a merger event (see Figure \ref{fig:outerburst} and discussion in section \ref{sec:results:perturbative}).
\citet{Buck2023} found similar predictions in the NIHAO-UHD simulations.
In our model, Range([Fe/H]) increases by $\sim$$0.1 - 0.2$ dex at a time coincident with the enhanced accretion associated with this model.
We therefore support their argument that the merger event with \gaia-Sausage Enceladus \citep[e.g.,][]{Belokurov2018, Helmi2018} should have had these effects on the MW disk.
We do however raise concerns regarding the reliability of birth radii determined with the current version of this methodology.
In both \citet{Lu2022b} and \citet{Ratcliffe2023}, Range([Fe/H]) and $\grad{Fe}$ as functions of age have the same but inverted shapes.
Figure \ref{fig:lu-comparison} indicates that a portion of the increase in MDF widths should be attributed to migration alone, and therefore Range([Fe/H]) and $\grad{Fe}$ should vary slightly differently with age.
The differences between the two may be quantifiable through a careful accounting of survey selection functions, which directly influence the measured radial migration strength and would introduce age-dependent effects.
\par
Lastly, we note that our GCE models predict MDFs that are systematically narrower than the data.
This mismatch is a shortcoming of our models whose origin is unclear.
One possibility is that the metallicity gradient was indeed steeper in the past, as argued by \citet{Lu2022b} and \citet{Ratcliffe2023}.
Another possibility is that radial migration is considerably faster than theoretical expectations such that stars sample a much broader range of the abundance gradient.
A third possibility, which is expected from hydrodynamic simulations \citep[e.g.,][]{Hopkins2014, Sparre2017, Feldmann2017, Ma2018}, is that departures from a smooth SFH on short timescales ($\lesssim$100 Myr) lead to variability in ISM abundances and therefore broader MDFs.
Such variations might complicate birth radius inferences with the \citet{Lu2022b} methodology.

\section{Conclusions}
\label{sec:conclusions}

In this paper, we advocate for a new explanation for the radial gradient in metal abundances of both gas and stars in MW-like spiral galaxies.
In this scenario, the ISM abundance evolves toward the local equilibrium abundance $Z_\text{eq}$ at all radii.
$Z_\text{eq}$ declines exponentially with radius with a scale length that is an intrinsic property of a galaxy set by GCE parameters and their variations with radius.
The evolution toward equilibrium is fast ($\sim$Gyr timescales), and the equilibrium state is reached early in the disk lifetime as a result.
In the absence of major mergers or changes in GCE parameters that disrupt the equilibrium, metal abundances do not evolve significantly once $Z_\text{eq}$ is reached, even with ongoing star formation.
Figure \ref{fig:cartoon} shows a cartoon of this behavior.
The result is that observed metallicity gradients closely trace the underlying equilibrium state for much of the disk lifetime.
\citet{Sharda2021a} found similar predictions using analytic models with a detailed treatment of gas dynamics.
\par
The equilibrium scenario is constrained first and foremost by the apparent age-independence of stellar metallicities across much of the MW disk.
We highlight this result in the \astronn\ value added catalog from SDSS-IV (see Figure \ref{fig:gradxh-fixedage}), though similar results have been a running theme in the literature in recent years.
\citet{Willett2023} demonstrated that conventional GCE models \citep[e.g.,][]{Chiappini2009, Minchev2013, Minchev2014} underpredict the metallicities of old stars with asteroseismic age estimates.
Using six different age catalogs, \citet{Gallart2024} showed that the relation between stellar age and metallicity is flat up to $\sim$10 Gyr in the solar neighborhood.
The metal abundances of open clusters \citep{Spina2022, Magrini2023, Carbajo-Hijarrubia2024} and classical Cepheid variables \citep{daSilva2023} do not correlate significantly with age across much of the Galactic disk.
\citet{Palla2024} modeled the open cluster abundances using recent metal-poor gas accretion.
Such an event is allowed by the equilibrium scenario.
In our models, major mergers result in perturbations that settle back to the equilibrium state on $\sim$Gyr timescales (see Figure \ref{fig:outerburst} and discussion in Section \ref{sec:results:perturbative}).
However, additional processes are required to hold the ISM metallicity constant on longer timescales ($\gtrsim$few Gyr), since re-enrichment following dilution events is fast \citep[e.g.,][]{Dalcanton2007, Johnson2020}.
\par
In our fiducial model (\modelname{Exp}{2}; see Table \ref{tab:gcemodels} and discussion in Section \ref{sec:gce:primary-set}), the equilibrium scenario arises from an increase in the outflow mass loading factor with Galactocentric radius (see Equation \ref{eq:zoeq-waf17} and discussion in sections \ref{sec:gce:primary-set} and \ref{sec:discussion:outflows}).
The outflow has two important effects.
First, it removes material from the ISM at the local abundance and replaces it with metal-poor gas through accretion, which lowers the equilibrium metallicity (see, e.g., the analytic models of \citealt{Weinberg2017b}).
Second, the outflow shortens the amount of time that baryons spend in the star forming ISM before they are either ejected or incorporated into new stars ($\lesssim$few Gyr; see Figs. \ref{fig:tauproc-demo} and \ref{fig:sfr-per-ifr} and discussion in Section \ref{sec:results:sfr-per-ifr}).
After a handful of ``generations'' of baryons, the ratio of star formation to accretion, $\dot{\Sigma}_\star / \dot{\Sigma}_\text{in}$, becomes nearly constant.
This quantity is closely related to the local equilibrium abundance, since metal production is most sensitive to the SFR while accretion sets the rate at which fresh H is introduced to the ISM (see Equation \ref{eq:zoeq-sfr-per-ifr}).
Changes in the relative balance of $\dot{\Sigma}_\star$ and $\dot{\Sigma}_\text{in}$, which are driven by changes in the outflow efficiency, set the equilibrium gradient slope $\nabla_\text{eq}$ (see Figures \ref{fig:sfr-ifr-gradslope} and \ref{eq:eq-gradient}).
We plan to explore radial gas flows \citep[e.g.,][]{Lacey1985, Bilitewski2012} as a potential alternative origin of the equilibrium scenario in future work (see also discussion in Section \ref{sec:discussion:outflows}).
\par
Our \modelname{Exp}{2} model is also quite similar to our fiducial model from \citetalias{Johnson2021} (see discussion in section 2 therein).
In that paper, we simply asserted the equilibrium scenario in order to isolate the effect of radial migration on the disk abundance structure for different assumptions of the SFH.
This paper offers empirical justification of that choice.
We demonstrated in \citetalias{Johnson2021} that this class of models readily explains many details of the disk-abundance structure, namely the dependence of the [O/Fe]-[Fe/H] distribution on Galactocentric radius and mid-plane distance \citep[e.g.,][]{Hayden2015, Vincenzo2021a} and the shapes of the age-[O/H], age-[Fe/H], and age-[O/Fe] relations \citep[e.g.,][]{Feuillet2018, Feuillet2019}.
The most noteworthy shortcoming of these models is that they fail to reproduce the distinct bimodality in [O/Fe] at fixed [Fe/H] (e.g., \citealt{Fuhrmann1998, Bensby2003, Adibekyan2012, Hayden2015}; see discussion in \citealt{Dubay2024}).
Many authors have proposed early accretion events and/or episodic SFHs in the early disk as explanations of this result (e.g., \citealt{Chiappini1997, Haywood2016, Mackereth2018, Spitoni2019, Buck2020b, Khoperskov2021}; Beane et al. 2024, in preparation).
Our models do not include such events.
We plan to explore these evolutionary histories in future work.
\par
% Using analytic models with a careful treatment of gas dynamics, \citet{Sharda2021a} also predict galaxy metallicity gradients to be in an equilibrium state across a broad range of redshift.
% We have arrived at the same conclusion using GCE models compared to ages and abundances of thin disk stars in the MW.
% Our combined results would suggest that radial metallicity gradients arise an equilibrium phenomenon driven by gas flows within the disk, between the disk and its surroundings, or some combination of the two.
% We plan to explore the effects of radial gas flows \citep[e.g.,][]{Lacey1985, Bilitewski2012} in GCE models in future work to investigate this hypothesis further.
% \par
The equilibrium scenario also offers a simple explanation for the apparent uniformity of metallicity gradients in galaxy disks.
Population studies have found a relatively tight Gaussian distribution of slopes \citep[e.g.,][]{Ho2015, Sanchez2020}.
There is no clear correlation with galaxy morphology, including the presence of a bar \citep{Zaritsky1994, Sanchez-Menguiano2016}.
Some galaxies have gradients that invert in the innermost regions \citep[e.g.,][]{Rosales-Ortega2011, Sanchez2014}, flatten in the outermost regions \citep[e.g.,][]{Martin1995, Bresolin2012}, or both \citep[e.g.,][]{Sanchez-Menguiano2018}.
Some authors argue that the slope is weakly correlated with galaxy stellar mass \citep[e.g.,][]{Perez-Montero2016, Belfiore2017, Groves2023}, while others do not find a clear a trend \citep[e.g.,][]{Bresolin2019, Pilyugin2019, Poetrodjojo2018, Poetrodjojo2021}.
At least for MW-like galaxies within the radial range of interest in this paper, there is a striking level of uniformity when scaled for disk size (see also Figure 9 of \citealt{Berg2020} and discussion in their Section 4.3).
These results suggest that whatever mechanism sets radial metallicity gradients is likely relatively universal among galaxies, such as a preferred equilibrium state.
% If metallicity gradients are set by differences in the star formation and assembly timescales between the inner and outer regions (i.e. inside-out growth; e.g., \citealt{Bird2013}), then one would perhaps expect greater diversity among the galaxy population due to variations in assembly histories.
% These results suggest that whatever sets radial metallicity gradients is likely relatively universal between galaxies, such a preferred equilibrium state.
\par
In the near future, we will be able to test the equilibrium scenario further with larger samples of stars provided by SDSS-V \citep{Kollmeier2017}.
Subgiants may be numerous enough in the available data to provide more direct constraints using isochrone ages \citep[e.g.,][]{Pont2004, Jorgensen2005}.
As large samples with increasingly precise age and abundance measurements become available, our understanding of the assembly and enrichment histories of the MW will improve accordingly.

\section*{Acknowledgements}
\label{sec:acknowledgements}

We are grateful to Joshua Simon, Andrew McWilliam, and Fiorenzo Vincenzo for valuable discussions.
Much of the discussion in this manuscript was inspired by conversations with various conference and seminar attendees, including Marco Palla, Emanuele Spitoni, Alexander Stone-Martinez, Kathryn Kreckel, Laura S{\'a}nchez-Menguiano, Diane Feuillet, Sofia Feltzing, Oscar Agertz, Christian Lehmann, Heitor Ernandes, Michael Blanton, David Hogg, Ivan Minchev, Jon Holtzman, Keith Hawkins, and Ralph Sch{\"o}nrich.
\par
This work was supported in part by National Science Foundation grant AST-2307621.
JWJ acknowledges financial support from an Ohio State University Presidential Fellowship and a Carnegie Theoretical Astrophysics Center postdoctoral fellowship.
GAB acknowledges the support from the ANID Basal project FB210003.
EJG is supported by a National Science Foundation Astronomy and Astrophysics Postdoctoral Fellowship under award AST-2202135.
\par
This research was supported by the Munich Institute for Astro-, Particle and BioPhysics (MIAPbP) which is funded by the Deutsche Forschungsgemeinschaft (DFG, German Research Foundation) under Germany's Excellence Strategy – EXC-2094 – 390783311.
\par
Funding for the Sloan Digital Sky Survey V has been provided by the Alfred P. Sloan Foundation, the Heising-Simons Foundation, the National Science Foundation, and the Participating Institutions. SDSS acknowledges support and resources from the Center for High-Performance Computing at the University of Utah. SDSS telescopes are located at Apache Point Observatory, funded by the Astrophysical Research Consortium and operated by New Mexico State University, and at Las Campanas Observatory, operated by the Carnegie Institution for Science. The SDSS web site is www.sdss.org.
\par
SDSS is managed by the Astrophysical Research Consortium for the Participating Institutions of the SDSS Collaboration, including Caltech, the Carnegie Institution for Science, Chilean National Time Allocation Committee (CNTAC) ratified researchers, The Flatiron Institute, the Gotham Participation Group, Harvard University, Heidelberg University, The Johns Hopkins University, L’Ecole polytechnique fédérale de Lausanne (EPFL), Leibniz-Institut für Astrophysik Potsdam (AIP), Max-Planck-Institut für Astronomie (MPIA Heidelberg), Max-Planck-Institut für Extraterrestrische Physik (MPE), Nanjing University, National Astronomical Observatories of China (NAOC), New Mexico State University, The Ohio State University, Pennsylvania State University, Smithsonian Astrophysical Observatory, Space Telescope Science Institute (STScI), the Stellar Astrophysics Participation Group, Universidad Nacional Autónoma de México, University of Arizona, University of Colorado Boulder, University of Illinois at Urbana-Champaign, University of Toronto, University of Utah, University of Virginia, Yale University, and Yunnan University.
\par
{\it Software}: {\sc NumPy} \citep{Harris2020}, {\sc Matplotlib} \citep{Hunter2007}, {\sc SciPy} \citep{Virtanen2020}

% \newpage
% \clearpage
\appendix

\section{The Leung et al. (2023) age catalog}
\label{sec:leung2023}

\begin{figure*}
\centering
\includegraphics[scale = 0.9]{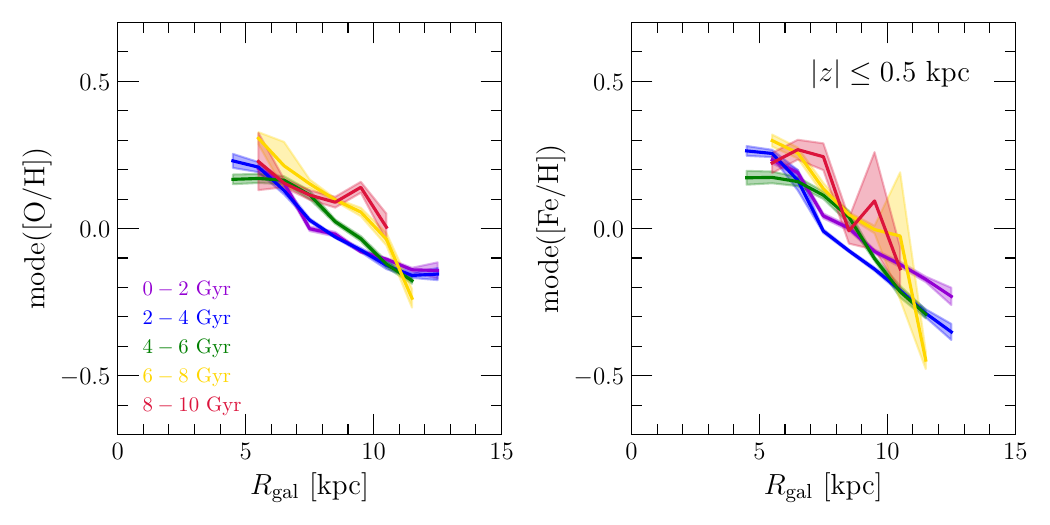}
\includegraphics[scale = 0.85]{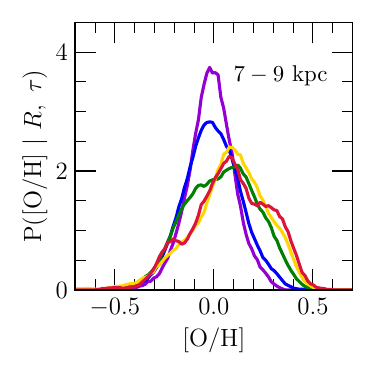}
\includegraphics[scale = 0.87]{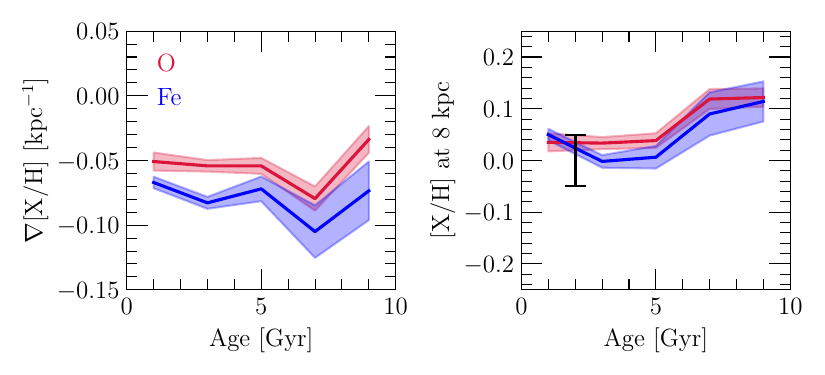}
\caption{
The same as Figure \ref{fig:gradxh-fixedage}, but using \citeauthor{Leung2023}'s \citeyearpar{Leung2023} catalog of stellar ages as opposed to the \astronn\ value added catalog.
Because of the smaller sample, we use 2 Gyr bins in age as opposed to 1 Gyr according to the legend in the top left panel.
{\bf Summary}: We find the same results as in Figure \ref{fig:gradxh-fixedage}, so our central results should not be affected by any learned correlations between stellar abundances and ages in the \astronn\ catalog.
}
\label{fig:latentage}
\end{figure*}

\begin{table}
\caption{
A summary of linear regression parameters applied to the [O/H]-$R$ and [Fe/H]-$R$ relations for mono-age populations in the \citet{Leung2023} catalog (see discussion in Appendix \ref{sec:leung2023}).
}
\begin{tabularx}{\columnwidth}{c @{\extracolsep{\fill}} c c c c}
% \toprule
\hline
Age Range & $\grad{O}$ & [O/H](R = 8 kpc) & $\grad{Fe}$ & [Fe/H](R = 8 kpc)
\\
\hline
% \\[-5pt]
% \multicolumn{3}{c}{\textbf{[O/H]}}
% \\[5pt]
$0 - 2$ Gyr & $-0.051 \pm 0.007$ kpc$^{-1}$ & $0.035 \pm 0.017$ & $-0.067 \pm 0.004$ kpc$^{-1}$ & $0.050 \pm 0.011$
\\
$2 - 4$ Gyr & $-0.054 \pm 0.004$ kpc$^{-1}$ & $0.034 \pm 0.012$ & $-0.083 \pm 0.005$ kpc$^{-1}$ & $-0.002 \pm 0.012$
\\
$4 - 6$ Gyr & $-0.054 \pm 0.006$ kpc$^{-1}$ & $0.038 \pm 0.014$ & $-0.072 \pm 0.010$ kpc$^{-1}$ & $0.006 \pm 0.022$
\\
$6 - 8$ Gyr & $-0.079 \pm 0.009$ kpc$^{-1}$ & $0.119 \pm 0.019$ & $-0.105 \pm 0.020$ kpc$^{-1}$ & $0.090 \pm 0.042$
\\
$8 - 10$ Gyr & $-0.034 \pm 0.010$ kpc$^{-1}$ & $0.122 \pm 0.018$ & $-0.073 \pm 0.023$ kpc$^{-1}$ & $0.114 \pm 0.039$
\\
\hline
% $0 - 2$ Gyr & $-0.067 \pm 0.004$ kpc$^{-1}$ & $0.050 \pm 0.011$
% \\
% $2 - 4$ Gyr & $-0.083 \pm 0.005$ kpc$^{-1}$ & $-0.002 \pm 0.012$
% \\
% $4 - 6$ Gyr & $-0.072 \pm 0.010$ kpc$^{-1}$ & $0.006 \pm 0.022$
% \\
% $6 - 8$ Gyr & $-0.105 \pm 0.020$ kpc$^{-1}$ & $0.090 \pm 0.042$
% \\
% $8 - 10$ Gyr & $-0.073 \pm 0.023$ kpc$^{-1}$ & $0.114 \pm 0.039$
% \\
% \hline
\end{tabularx}
\label{tab:latentage-regressions}
\end{table}

In this appendix, we replicate our results in Figure \ref{fig:gradxh-fixedage} with the \citet{Leung2023} catalog of stellar ages.
We have focused our discussion in this paper on the \astronn\ catalog \citep{Mackereth2019b}, but it is possible that their convolutional neural network simply learned the correlations between chemical abundances and ages.
In this case, our key empirical result in Figure \ref{fig:gradxh-fixedage} would be subject to considerable systematic uncertainties.
\citet{Leung2023} demonstrate that their estimates, which are also for APOGEE stars, do not contain any significant amount of information on the abundances of alpha and iron-peak elements (see discussion in Section \ref{sec:empirical:caveats}).
Their catalog is therefore an excellent comparison case to validate our measurements in Figure \ref{fig:gradxh-fixedage}.
\par
\citet{Leung2023} mitigate this potential issue by compressing the spectra into lower dimensional representations of themselves (i.e., a~\textit{latent space}) using a variational encoder-decoder algorithm~\citep[e.g.,][]{LeCun2015}.
They then train a modified random forest algorithm to predict similarly compressed lightcurves trained on~\textit{Kepler} photometry~\citep{Borucki2010}.
They demonstrate that this latent space contains little if any information on alpha and iron-peak element abundances as well as stellar parameters, as intended.
Red giant ages estimated through machine learning algorithms trained on spectra are much more strongly correlated with their C and N abundances \citep{Stone-Martinez2024}.
\par
The \citet{Leung2023} training sample spanned a much narrower range in surface gravity ($\log g = 2.5 - 3.6$), which lowers our sample size by a factor of $\sim$$2.5$.
We therefore use 2 Gyr as opposed to 1 Gyr bins in age and compute the mode for subsamples containing at least 100 as opposed to 200 stars, otherwise following the same procedure described in Section \ref{sec:empirical:evolution}.
Figure \ref{fig:latentage} shows the results, with the lines of best fit to the metallicity gradients in each age bin reported in Table \ref{tab:latentage-regressions}.
Consistent with our results in Figure \ref{fig:gradxh-fixedage}, we find that both the slope and normalization of the disk metallicity gradient are independent of age up to $\sim$10 Gyr, which is the full range of ages that we probe here.
We therefore conclude that our central results are not affected by any learned correlations between chemical abundances and stellar ages in the \astronn\ catalog.

\section{GCE Model Parameter Calibrations}
\label{sec:calibration}

In this appendix, we describe our procedure for assigning $\tau_\text{rise}$ and $\tau_\text{sfh}$ as functions of Galactocentric radius in our GCE models.
These parameters describe the shape of the SFH in a given ring (see Equation \ref{eq:rise-fall-sfh}).
Our procedure follows the analytic one-zone GCE models by \citet{Weinberg2017b}.
\par
In a given annulus within the MW disk, the time derivative of the gas surface density\footnote{
    The areas of the rings in our multi-zone models are constant, so surface density follows the same continuity equations as mass.
} follows a summation of source and sink terms:
\begin{equation}\begin{split}
\dot{\Sigma}_g &= \dot{\Sigma}_\text{in} - \dot{\Sigma}_\star - \dot{\Sigma}_\text{out} + \dot{\Sigma}_r
\\
&\approx \dot{\Sigma}_\text{in} - \dot{\Sigma}_\star
\left(1 + \eta - r\right),
\label{eq:dot-sigma-gas}
\end{split}\end{equation}
where $\dot{\Sigma}_r \approx r \dot{\Sigma}_\star$ is the rate of return of stellar envelopes back to the ISM (see discussion in Section \ref{sec:gce:primary-set}).
Our multi-zone models also follow Equation \refp{eq:dot-sigma-gas} under the caveat that stellar populations may be exchanged between rings, so the exact rate of return varies a small amount with time.
In fact, this stochastic exchange of mass is the reason for the ``jitter'' in the predicted accretion rates seen in Figs. \ref{fig:allmodels-evol}, \ref{fig:sfr-per-ifr}, and \ref{fig:sfr-ifr-gradslope}.
\par
The rate of change in the surface density of O is given by equation 7 of \citet{Weinberg2017b}:
\begin{equation}
\dot{\Sigma}_\text{O} = \ycc{O} \dot{\Sigma}_\star -
Z_\text{O} \dot{\Sigma}_\star
\left(1 + \eta - r\right).
\label{eq:dot-sigma-o}
\end{equation}
There is an additional factor of $Z_\text{O}$ in the second term, because losses of O to star formation and outflows occur at the ISM abundance at a given time.
The source term $\ycc{O} \dot{\Sigma}_\star$ describes CCSN production of O under the approximation of instantaneous production, which is accurate enough for our purposes due to the short lifetimes of massive stars \citep[e.g.,][]{Larson1974, Maeder1989, Henry2000}.
The rate of change of the O abundance $\dot{Z}_\text{O}$ then follows from combining equations \ref{eq:dot-sigma-gas} and \ref{eq:dot-sigma-o} with quotient rule:
\begin{equation}\begin{split}
\dot{Z}_\text{O} &= \frac{
    \Sigma_g \dot{\Sigma}_\text{O} - \Sigma_\text{O} \dot{\Sigma}_g
}{
    \Sigma_g^2
}
\\
&= \frac{\ycc{O}}{\tau_\star} - \frac{Z_\text{O}}{\tau_\star}
\left(
1 + \eta - r + \tau_\star \frac{\dot{\Sigma}_g}{\Sigma_g}
\right),
\label{eq:dot-z-o}
\end{split}\end{equation}
where we have also substitued in the SFE timescale $\tau_\star \equiv \Sigma_g / \dot{\Sigma}_\star$.
\par
We hold $\tau_\star$ constant in time for the purposes of this parameter calibration.
In our numerical models, we use the full time-evolution described by Equation \refp{eq:tau-star}.
Under this assumption, the rates of change in the SFR and the gas supply are related by $\ddot{\Sigma}_\star / \dot{\Sigma}_\star = \dot{\Sigma}_g / \Sigma_g$.
Differentiating our rise-fall SFH (see Equation \ref{eq:rise-fall-sfh}) with time and plugging in the result yields the following expression for $\dot{Z}_\text{O}$:
\begin{equation}
\dot{Z}_\text{O} = \frac{\ycc{O}}{\tau_\star} -
\frac{Z_\text{O}}{\tau_\star} \left(
1 + \eta - r - \frac{
    \tau_\star e^{-t / \tau_\text{rise}}
}{
    \tau_\text{rise} \left(1 - e^{-t / \tau_\text{rise}}\right)
} - \frac{\tau_\star}{\tau_\text{sfh}}
\right).
\end{equation}
This expression is a linear ordinary differential equation, whose solution has a known form (see equation 33 of \citealt{Weinberg2017b}).
In this case, the solution is given by
\begin{equation}
Z_\text{O}(t) = Z_\text{O,eq} \left[
1 - \exp\left(-t \frac{
    \tau_\text{sfh} - \tau_\text{proc}
}{
    \tau_\text{sfh}\tau_\text{proc}
}\right) - \frac{
    \tau_\text{sfh}\tau_\text{rise}
}{
    \tau_\text{sfh}\tau_\text{rise} -
    \tau_\text{proc}\tau_\text{rise} -
    \tau_\text{sfh}\tau_\text{proc}
}\left(
e^{-t / \tau_\text{rise}} - \exp\left(
-t \frac{
    \tau_\text{sfh} - \tau_\text{proc}
}{
    \tau_\text{sfh}\tau_\text{proc}
}
\right)
\right)
\right],
\label{eq:zo-full-soln}
\end{equation}
where $Z_\text{O,eq}$ is the equilibrium abundance (see Equation \ref{eq:zoeq-waf17}), and $\tau_\text{proc}$ is the processing timescale (see Equation \ref{eq:tauproc}).
The integration constant is assigned such that $Z_\text{O} = 0$ at $t = 0$.

\begin{figure*}
\centering
\includegraphics[scale = 1]{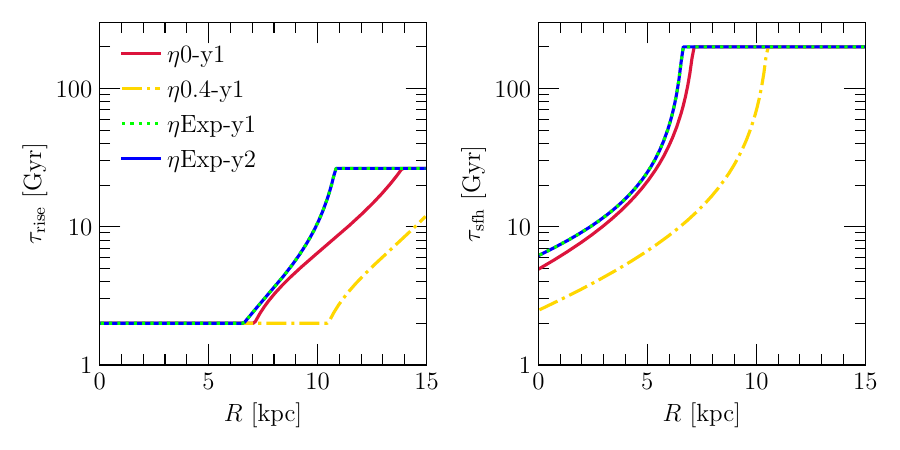}
\caption{Our SFH parameter calibration.
Curves show $\tau_\text{rise}$ (left) and $\tau_\text{sfh}$ (right) as a function of radius in each model, marked according to the legend in the left panel.
{\bf Summary}: Each model evolves under the inside-out paradigm in that the SFH becomes more extended with increasing Galactocentric radius. 
}
\label{fig:calibration}
\end{figure*}

In our \modelname{0}{1} and \modelname{0.4}{1} models, we assign each ring values of $\tau_\text{rise}$ and $\tau_\text{sfh}$ such that Equation \refp{eq:zo-full-soln} predicts the O abundance observed in the ISM at the present day.
Following Figure \ref{fig:gradients}, we use the measurements by \citet{MendezDelgado2022} as our empirical benchmark.
As described in Section \ref{sec:gce:calibration}, we first assume $\tau_\text{rise} = 2$ Gyr and search for values of $\tau_\text{sfh}$ between $0.1$ and $200$ Gyr satisfying these criteria.
If no solution is found, we adopt $\tau_\text{sfh} = 200$ Gyr and search for values of $\tau_\text{rise}$ between $2$ Gyr and $2\tau_\text{disk} = 26.4$ Gyr.
If still no solution is found, we simply adopt $\tau_\text{rise} = 26.4$ Gyr, which corresponds to an SFH that rises approximately linearly until the present day.
For the sake of this parameter calibration, we use our scaling of $\tau_\star$ with gas surface density at the present day given by Equation \refp{eq:tau-star} assuming a simple exponential disk with a scale radius of $R_g = 3.75$ kpc \citep{Kalberla2009} and $\tau_\star = 2$ Gyr at $R = 0$.
\par
In our \modelname{Exp}{1} and \modelname{Exp}{3} models, the present-day ISM metallicities are much more sensitive to the choice of mass loading factor $\eta$ than the shape of the SFH.
For these models, we therefore choose $\tau_\text{rise}$ and $\tau_\text{sfh}$ such that the 50th percentile of the integrated SFH matches the observed median age at a given radius:
\begin{equation}
\int_0^{\tau_{1/2}(R)} (1 - e^{-t / \tau_\text{rise}})
e^{-t / \tau_\text{sfh}} dt =
\frac{1}{2} \int_0^{\tau_\text{disk}}
(1 - e^{-t / \tau_\text{rise}}) e^{-t / \tau_\text{sfh}} dt.
\end{equation}
We otherwise follow the same procedure as described above, starting with $\tau_\text{rise} = 2$ Gyr and searching for a value of $\tau_\text{sfh}$ between $0.1$ and $200$ Gyr.
\par
Figure \ref{fig:calibration} shows the resulting values of $\tau_\text{rise}$ and $\tau_\text{sfh}$ as a function of radius in our primary set models.
The \modelname{0.4}{1} model finds a solution at all radii, while the \modelname{0}{1} model resorts to the ``fail-safe'' values described above only at $R \gtrsim 13.5$ kpc.
Our \modelname{Exp}{1} and \modelname{Exp}{2} models resort to these values at $R \gtrsim 10.5$ kpc.
However, this shortcoming of our parameter selection does not affect our main conclusions.
The median stellar ages predicted by these models are in broad agreement with the observations at these radii anyway (see the bottom-left panel of Figure \ref{fig:gradients} and the right panel of Figure \ref{fig:calibration-results}).
Our primary interest also lies not with the age gradients but with the metallicity gradients predicted by these models, which are much more sensitive to the mass loading factor $\eta$.

\bibliographystyle{aasjournal}
\bibliography{main}

\end{document}